\documentclass[preprint]{aastex62}

\pdfoutput=1

\submitjournal{ApJ}

\shorttitle{Black Hole X-ray Spectra from Simulations}
\shortauthors{Kinch et al.}

\begin{document}

\title{Predicting Stellar-Mass Black Hole X-ray Spectra from Simulations}

\correspondingauthor{Brooks E. Kinch}

\author[0000-0002-8676-425X]{Brooks E. Kinch}
\affil{Department of Physics and Astronomy, Johns Hopkins University,
Baltimore, MD 21228, USA}
\email{kinch@jhu.edu}

\author{Jeremy D. Schnittman}
\affil{NASA Goddard Space Flight Center,
Greenbelt, MD 20771, USA}
\email{jeremy.d.schnittman@nasa.gov}

\author{Timothy R. Kallman}
\affil{NASA Goddard Space Flight Center,
Greenbelt, MD 20771, USA}
\email{timothy.r.kallman@nasa.gov}

\author{Julian H. Krolik}
\affil{Department of Physics and Astronomy, Johns Hopkins University,
Baltimore, MD 21228, USA}
\email{jhk@jhu.edu}

\begin{abstract}
We describe results from a new technique for the prediction of complete, self-consistent X-ray spectra from three-dimensional General Relativistic magnetohydrodynamic (GRMHD) simulations of black hole accretion flows. Density and cooling rate data from a \textsc{harm3d} GRMHD simulation are processed by both an improved version of the Monte Carlo radiation transport code \textsc{pandurata} (in the corona) and the Feautrier solver \textsc{ptransx} (in the disk), with \textsc{xstar} subroutines. The codes are run in a sequential but iterative fashion to achieve globally energy-conserving and self-consistent radiation fields, temperature maps, and photoionization equilibria. The output is the X-ray spectrum as seen by a distant observer. For the example cases we consider here---a non-rotating $10 M_\odot$ black hole with solar abundances, accreting at 0.01, 0.03, 0.1, or 0.3 Eddington---we find spectra resembling actual observations of stellar-mass black holes in the soft or steep power-law state: broad thermal peaks (at 1-3 keV), steep power-laws extending to high energy ($\Gamma = 2.7$-4.5), and prominent, asymmetric Fe K$\alpha$ emission lines with equivalent widths in the range 40-400 eV (larger EW at lower accretion rates). By starting with simulation data, we obviate the need for parameterized descriptions of the accretion flow geometry---no \emph{a priori} specification of the corona's shape or flux, or the disk temperature or density, etc., are needed. Instead, we apply the relevant physical principles to simulation output using appropriate numerical techniques; this procedure allows us to calculate inclination-dependent spectra after choosing only a small number of physically meaningful parameters: black hole mass and spin, accretion rate, and elemental abundances.
\end{abstract}

\keywords{accretion, accretion disks --- black hole physics --- line: formation}

\section{Introduction}

Accreting black holes provide a unique opportunity to investigate both the strong field regime of General Relativity and the physics of accretion flows in extreme environments. Both active galactic nuclei and stellar-mass black hole binaries produce X-ray spectra with line and continuum features which convey information about the environment and spacetime geometry from which they originate. Relativistically-broadened Fe K$\alpha$ fluorescence lines are one of the key indicators that these systems do in fact contain black holes \citep{tan95a}, and the thermal plus power-law continuum indicates the presence of disk and corona, respectively \citep{lia79a, haa91a}. Indeed, studying the governing physics of accretion processes is tied to our ability to connect the underlying theory to observations. The quantitative information inferrable from any spectrum is, however, limited by the templates to which the observation is compared.

To this end, we have developed a technique with which model spectra are computed directly from simulation data by applying the relevant physical principles while invoking almost no assumptions. The numerical machinery we describe here is an extension of that introduced in \citet{kin16a}; as in that paper, we apply our method only to the case of non-rotating, stellar-mass black holes, but we otherwise greatly expand the predictive scope of our method by treating the X-ray emission lines \emph{and} the continuum in a self-consistent, energy-conserving fashion. In addition, we explore the effects of varying the nominal accretion rate on the predicted spectrum.

A variety of methods are currently in use for modeling the spectra of black hole systems. Some are purely phenomenological---the continuum is fit with a multicolor disk blackbody (e.g., \textsc{diskbb} \citep{mit84a}) plus a (typically broken) power-law at high energy, and the Fe K$\alpha$ emission from the disk surface is assumed to vary as a decreasing power-law (or sometimes broken power-law) in radius with a hard cutoff at the innermost stable circular orbit (ISCO) and another at some outer radius (e.g., \textsc{relline} \citep{dau13a}; see \cite{rey03a} for a discussion of these methods). More sophisticated techniques model the continuum with a single-zone Comptonization region and the disk reprocessed component (the Fe K$\alpha$ line, the K-edge, and the Compton bump) by performing detailed radiative transfer and photoionization calculations within a sample section of the disk (e.g., the codes \textsc{reflionx} \citep{ros05a}, \textsc{xillver} \citep{gar10a, gar11a, gar13a}, and \textsc{relxill} \citep{gar14a}). At present \emph{all} methods rely in some way upon a parameterized description of the black hole environment: at best, an idealized corona (often a ``lamppost'' point source or a single homogeneous region) emits a power-law spectrum (perhaps with a thermal cutoff) which illuminates a semi-infinite, blackbody-radiating disk, and this disk has a knife-edge cutoff precisely at the ISCO. When using such a model to, for example, extract spin measurements from spectral data \citep{rey13a, mil15a}, the accuracy of the measurement is limited by the accuracy of the assumed accretion flow geometry and associated coronal flux.

By starting with 3D GRMHD simulation data, we greatly reduce the number of assumptions needed to describe the accretion flow geometry. We therefore also reduce the number of free parameters needed to specify a resulting observable spectrum. We do not, for example, require a sharp cutoff in Fe K$\alpha$ emission at the ISCO, nor do we specify the coronal geometry (lamppost or otherwise) \emph{a priori}; the density and temperature structure of the disk are not assumed in advance either. Instead, these are computed directly from the underlying physics, with the sole significant assumption being the equation of state employed by the simulation (as we describe below); as simulations improve, e.g., by the use of more realistic equations of state, our apparatus can easily be applied to their output as well. The resulting prediction of our method, the full inclination-dependent observable spectrum, is a function of a very small number of parameters, each physically meaningful: the black hole mass and spin, the nominal accretion rate, and the elemental abundances.

\section{Method}

\begin{figure}
\plotone{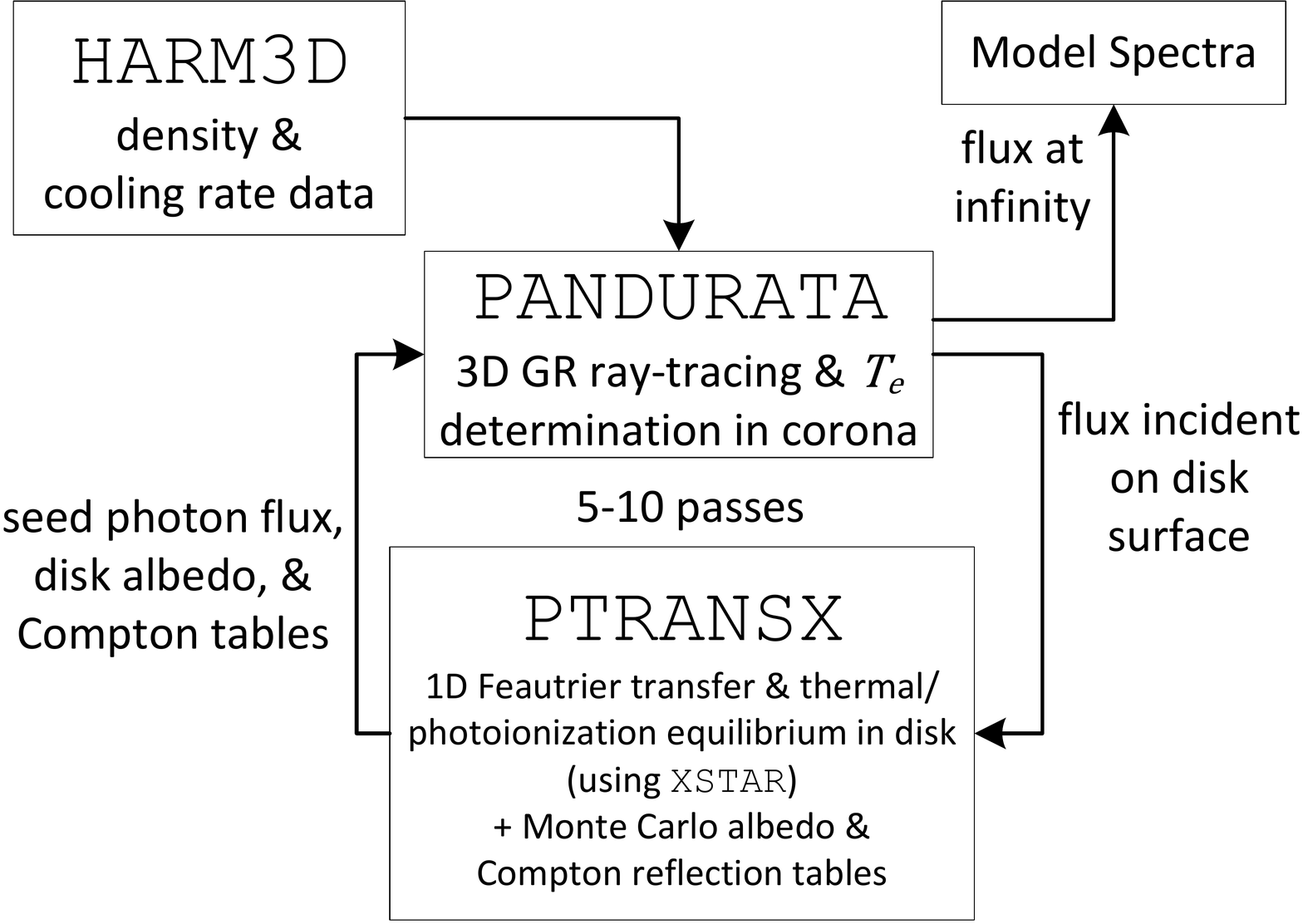}
\caption{A schematic overview of the general procedure. \label{fig:flowchart}}
\end{figure}

Our procedure has three main components. First, an accreting black hole system is simulated using \textsc{harm3d} \citep{nob09a}. We take a three-dimensional snapshot of the fluid density, four-velocity, and dissipation (cooling) rate at a time when the simulation has achieved approximate inflow equilibrium (out to $r \sim 20 M$). Using a thermal seed photon injection rate computed by integrating the local dissipation rate within the disk's photosphere, the Monte Carlo radiation transport code \textsc{pandurata} \citep{sch13b} determines the radiation field consistent with the simulation data and thermal balance in the corona \citep{sch13a}. With \textsc{harm3d}'s description of the disk structure and \textsc{pandurata}'s calculation of the disk incident flux, \textsc{ptransx} computes the disk's reprocessed \emph{outgoing} flux, requiring photoionization equilibrium and energy conservation everywhere within the disk \citep{kin16a}. This step yields a new guess for both the energy-dependent seed photon flux emerging from the disk surface and the spatial- and energy-dependent disk albedo---input for the next \textsc{pandurata} run. We cycle between \textsc{pandurata} (in the corona) and \textsc{ptransx} (in the disk) until a consistent picture of the \emph{global} radiation field develops. This cycling is a significant improvement to the original \textsc{pandurata} method. To avoid confusion between the iterative procedures within each step and the outermost iterations between \textsc{pandurata} and \textsc{ptransx}, we refer to the latter as ``passes.'' With one final round of relativistic ray-tracing, both reprocessed disk and coronal emission are transported to a distant observer in order to construct the complete predicted spectrum. The overall scheme is summarized in Figure \ref{fig:flowchart}.

\subsection{Simulation Data --- \textsc{harm3d}}
\label{sim_data}

The density ($\rho$ or $n_e$) and cooling rate ($\mathcal{L}$) data are from one snapshot of a \textsc{harm3d} simulation, taken at a time when the disk is in statistically steady state. \textsc{harm3d} is a three-dimensional, intrinsically conservative General Relativistic Magnetohydrodynamic (GRMHD) code, with a cooling function designed to produce a geometrically thin disk. \textsc{harm3d} solves a modified stress-energy conservation equation: in gravitationally-bound gas above a target temperature $T_*$, the excess heat is radiated away on an orbital timescale; $T_*$ is chosen so as to achieve a target aspect ratio \citep{nob09a}. The specific simulation we use, ``ThinHR'' \citep{nob11a}, has an aspect ratio $H_{\text{dens}}/r = 0.06$ (where $H_{\text{dens}}$ is the density-weighted scale height), and is still one of the best-resolved GRMHD disk simulations ever carried out \citep{haw13a}.

Translating the simulation data from ``code units'' to physical (cgs) units requires specification of the central black hole mass $M$, which sets the length and time scales ($ 1 M = (M/M_\odot) \cdot 1.5 \times 10^5 \ \text{cm} = (M/M_\odot) \cdot 4.9 \times 10^{-6} \ \text{s}$), and the accretion rate (in Eddington units) $\dot{m}$, which sets the scale for the density and cooling rate through \citep{sch13a}:
\begin{equation}
\rho_{\text{cgs}} = \rho_{\text{code}} \frac{4 \pi c^2}{\kappa G M} \frac{\dot{m} / \eta}{\dot{M}_{\text{code}}},
\label{eq:rho}
\end{equation}
\begin{equation}
\mathcal{L}_{\text{cgs}} = \mathcal{L}_{\text{code}} \frac{4 \pi c^7}{\kappa G^2 M^2} \frac{\dot{m} / \eta}{\dot{M}_{\text{code}}},
\label{eq:L}
\end{equation}
where $\kappa = 0.4 \ \text{cm}^2 \ \text{g}^{-1}$ is the electron scattering opacity and $\eta = 0.061$ ($> 0.057$, the \citet{nov73a} value) is the radiative efficiency found in that simulation \citep{nob11a}. In this paper, we consider a $10 M_\odot$ central black hole at four accretion rates, $\dot{m} = 0.01$, 0.03, 0.1, and 0.3.

\begin{figure}
\plotone{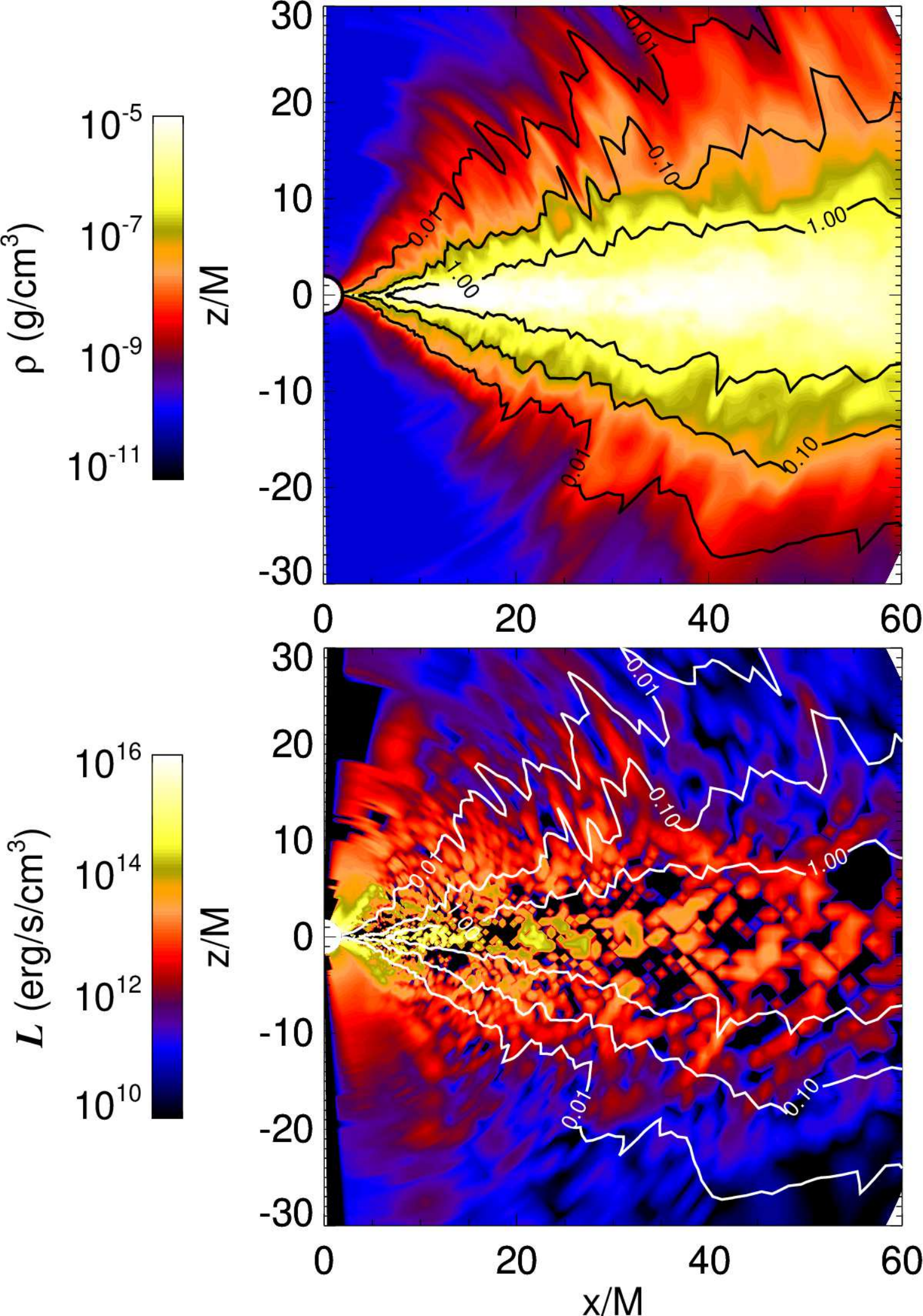}
\caption{The $x-z$ slice of the \textsc{harm3d} simulation data we use in this paper, scaled to $M = 10 M_\odot$, $\dot{m} = 0.01$, with three surfaces of constant optical depth overlaid. Note the great difference (and rapid change) in density between the disk body and the corona (upper panel). The lower panel shows the local instantaneous cooling rate---not all fluid elements are radiatively cooling at each time step (see \citet{sch13a} for a more extensive discussion of the cooling function). There is significant cooling in the corona, even where the density is very low. \label{fig:harm_slice}}
\end{figure}

With a known density structure (and a known spacetime geometry), surfaces of constant optical depth can be defined by integrating the electron scattering opacity along arcs of constant $(r, \phi)$, starting from the poles and continuing until the desired optical depth is reached. With the disk lying in the $x-y$ plane, the collection of points $\Theta(r, \phi)$ which satisfy
\begin{equation}
\int_0^{\Theta(r, \phi)} \sigma_T n_e \sqrt{g_{\theta\theta}} d\theta = \tau \ \mathrm{or}\ \int_{\Theta(r, \phi)}^\pi \sigma_T n_e \sqrt{g_{\theta\theta}} d\theta = \tau
\end{equation}
(where $\theta$ is the polar angle) defines the (upper and lower) surfaces of constant optical depth for the given value of $\tau$. The natural choice of surfaces with which to divide the disk body from the corona are the $\tau = 1$ surfaces, which we call the disk photospheres and label $\Theta_{\text{top}}$ and $\Theta_{\text{bot}}$. At any given $(r, \phi)$, the region between these surfaces---if they exist---is the disk body; everywhere else is the corona. Which $\tau$ value to use for dividing the disk and corona is somewhat arbitrary. For our purposes, a division is needed such that the only significant cooling process in the corona is inverse Compton (IC) scattering, while all atomic processes (such as Fe K$\alpha$ production) occur within the disk. Because the maximum local ratio of free-free power to net IC power we compute \emph{post hoc} in the corona is $\lesssim 4\%$ (just outside $\tau = 1$), and we find significant Fe K$\alpha$ production limited typically to $\tau > 1.5$, the $\tau = 1$ surface is a satisfactory choice for the photosphere. Figure \ref{fig:harm_slice} shows a cross section of \textsc{harm3d} density and cooling data, scaled to $10 M_\odot$ and $\dot{m} = 0.01$, with several surfaces of constant optical depth overlaid.

\subsection{Coronal Radiation Field --- \textsc{pandurata}}

For the initial pass, we assume that the total cooling at one $(r, \phi)$ arc within the disk body is radiated thermally at its photospheres. That is, at any $(r, \phi)$ for which the $\tau = 1$ surfaces exist, the flux outward at both $\Theta_{\text{top}}$ and $\Theta_{\text{bot}}$ is described as a hardened blackbody with effective temperature
\begin{equation}
\int_{\Theta_{\text{top}}}^{\Theta_{\text{bot}}} \mathcal{L} \sqrt{g_{\theta\theta}} d\theta = 2 \sigma T_{\text{eff}}^4.
\end{equation}
These thermal seed photons are ray-traced through the corona by the Monte Carlo radiation transport and local temperature balance code \textsc{pandurata}. For all subsequent passes, \textsc{pandurata} uses \textsc{ptransx}'s output seed photon spectra instead of the hardened blackbody. \textsc{pandurata} takes as input the density and cooling maps from \textsc{harm3d}, as well as the seed photon emission at the disk photosphere, and outputs: (1) an electron temperature map of the corona; (2) the spectrum as seen by distant observers; and (3) the spectral shape and strength of the flux incident upon the (upper and lower) disk photospheres at each $(r, \phi)$.

\begin{figure}
\plotone{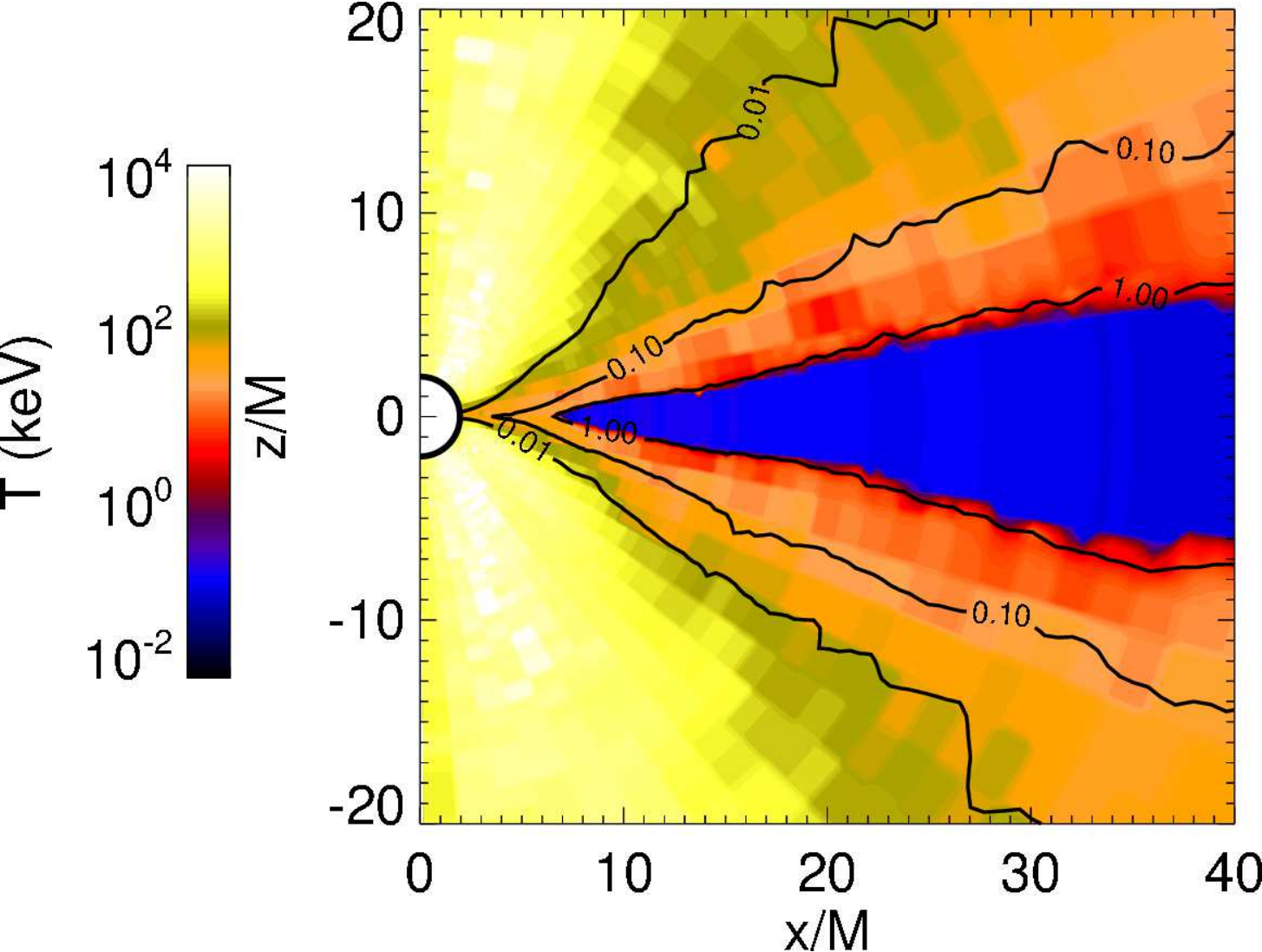}
\caption{The electron temperature map produced by \textsc{pandurata} for the same slice of data shown in Figure \ref{fig:harm_slice} ($M = 10 M_\odot$, $\dot{m} = 0.01$). Note that within the disk body (between the $\tau = 1$ surfaces) the electron temperature is shown as the constant value $T_{\text{eff}}$; \textsc{pandurata} does not determine the electron temperature within the disk body. The visible ``blockiness'' is due to our division of the coronal volume into sectors. \label{fig:t_map}}
\end{figure}

The operation of \textsc{pandurata}'s original version is described in detail in \citet{sch13b}. In brief, the code simulates the trajectories and scattering of seed photons in the corona while solving for the electron temperature at each point in the corona by setting the net inverse Compton power equal to \textsc{harm3d}'s local cooling rate. Several modifications to \textsc{pandurata} were made for its use in this project. First, photon packet scattering off of single electrons was replaced by an \emph{ensemble} approach---when a photon packet scatters in the corona, the photon packet's spectrum is redistributed according to an angle-averaged energy redistribution function described below (see section \ref{trans_sol}); its new direction, however, is determined as if it  were a single photon scattering off a single electron whose particular velocity was selected from the Maxwell-J\"{u}ttner distribution. Second, the coronal volume is divided into sectors---a coarser grouping, compared to the underlying simulation grid, of $\sim 100$ contiguous grid-cells each---with the interior of each sector treated as having a single temperature; net IC power is assessed for the sector as a whole, and a sector's temperature is adjusted by way of a Newton-Raphson method until its net IC power equals its total internal cooling rate. These two changes to \textsc{pandurata} (compared to its description in \citet{sch13b}) allow faster determination of the coronal temperature map, now necessary since \textsc{pandurata} must re-determine the temperature map each pass. We have verified that modified \textsc{pandurata} produces the same output spectrum as unmodified \textsc{pandurata}. The final modification, however, is more substantive: photon packets which strike the disk surface are subject to absorption and Compton recoil according to albedo and redistribution tables computed with \textsc{ptransx} output, using a procedure described in section \ref{repr_spec}. An example cross section of an electron temperature map so computed is shown in Figure \ref{fig:t_map}.

The end result is a complete description of the electron temperature and radiation field everywhere in the corona---including the flux irradiating the disk surface---that is consistent with the density and cooling structure of the GRMHD simulation as well as the temperature and ionization structure of the disk body.

\subsection{Disk Reprocessing --- \textsc{ptransx}}

\subsubsection{Defining the Problem}

At each $(r, \phi)$, the region between $\Theta_{\text{top}}$ and $\Theta_{\text{bot}}$ constitutes a column of the disk body. Its vertical density and cooling rate profiles are known from the \textsc{harm3d} simulation data, and the fluxes incident upon its upper and lower surfaces are computed by \textsc{pandurata}. In addition, some choice of elemental abundances is required. The chief assumption is that such a column can be treated as a finite, plane-parallel slab independent from its neighbors. The problem is to find, for each slab, a description of the radiation field and ionization balance at all vertical points that is energy-conserving, in photoionization equilibrium, consistent with the boundary conditions and structure provided, and which includes as many of the relevant physical processes as possible. Photoionization equilibrium is a reasonable assumption: for the densities and temperatures typical of the accretion disks we consider, the recombination timescale is short, $\sim 10^{-7}$ s (for highly-ionized Fe), compared to the disk's dynamical timescale, $\sim 10^{-3}$ s. We accomplish this with the code \textsc{ptransx}, a version of which was introduced in \citet{kin16a}; since then, \textsc{ptransx} has undergone major improvements and restructuring, particularly in its treatment of Compton scattering but in other areas as well. Its operation is described below.

\subsubsection{The Transfer Solution}
\label{trans_sol}

We solve the radiative transfer equation in plane-parallel geometry, including all relevant atomic processes and Compton scattering:
\begin{equation}
\mu \frac{d I_{\mu\varepsilon}}{d z} = -\alpha_\varepsilon I_{\mu\varepsilon} + j_\varepsilon - n_e \sigma_\varepsilon I_{\mu\varepsilon} + n_e \int_{-1}^1 \int_0^\infty d \mu^\prime d \varepsilon^\prime \sigma_{\varepsilon^\prime} R(\mu^\prime, \varepsilon^\prime; \mu, \varepsilon) I_{\mu^\prime \varepsilon^\prime}.
\end{equation}
We employ the Feautrier method \citep{mih78a}, which requires only that the redistribution function $R$ has forward-backward symmetry, i.e., that $R(\mu^\prime, \varepsilon^\prime; \mu, \varepsilon)$ is unchanged under
\begin{equation}
\Delta \theta = \cos^{-1} \left[\mu^\prime \mu + \sqrt{\left(1 - {\mu^\prime}^2\right)\left(1 - \mu^2\right)}\right] \to \Delta \theta \pm \pi.
\end{equation}
$R$ is a measure of the probability that photons with angle-energy $(\mu^\prime, \varepsilon^\prime)$ will scatter to angle-energy $(\mu, \varepsilon)$. With a specification of boundary conditions---$I_{\mu\varepsilon}$ \emph{inward} at the upper and lower surfaces (the incident intensity from \textsc{pandurata})---we solve a discretized version of the above transfer equation \emph{directly} via a forward-backward recursive sweep \citep{mih85a}.

Our treatment of Compton scattering is expressed by our choice of $R$. Though we have gone to great lengths to describe Compton scattering as accurately as possible, we are required by the Feautrier method to make the following approximation:
\begin{equation}
R(\mu^\prime, \varepsilon^\prime; \mu, \varepsilon) = \frac{1 + \left[\mu^\prime \mu + \sqrt{\left(1 - {\mu^\prime}^2\right)\left(1 - \mu^2\right)}\right]^2}{2} \mathcal{R} (\varepsilon^\prime, \varepsilon).
\end{equation}
We replace the angular dependence of a more accurate redistribution function with the dipole phase function of Thomson scattering, which has the required forward-backward symmetry. The Klein-Nishina cross section \emph{does not} have this symmetry---forward scattering is preferred to backward scattering, and significantly so at energies approaching and beyond $m_e c^2$. For the energies we are most concerned with ($\lesssim 50$ keV), however, that preference is modest. $\mathcal{R}$ is the angle-average of the full Compton redistribution function (itself a function of the local electron temperature) computed directly with an independent Monte Carlo calculation using relativistic dynamics, the Klein-Nishina cross section, and the Maxwell-J\"{u}ttner velocity distribution. The same $\mathcal{R}$ is used in the ensemble scattering calculation of \textsc{pandurata} described above.

\subsubsection{Equilibrium-Finding Procedure}

We make use of subroutines of the photoionization code \textsc{xstar} \citep{kal01a} in order to compute the local ionization balance---and consequent emissivity and absorption opacity---of gas at a fixed temperature and density, immersed in a known radiation field, in photoionization equilibrium.

At each $(r, \phi)$ sampled, the disk body is divided into some number of vertical cells (typically a few dozen, see section \ref{num_specs} below). For the $i^{\text{th}}$ cell, the net energy balance $y_i$ is defined as the difference between the net energy flux \emph{out} of the cell and the total cooling within that cell. That is:
\begin{equation}
y_i = \int_0^\infty d\varepsilon \left(F_{i,\varepsilon}^{\text{top}} - F_{i,\varepsilon}^{\text{bot}}\right) - \mathcal{L}_i (\Delta z)_i.
\label{eq:y}
\end{equation}
The collection of these values for all cells in the given vertical column forms the vector $\mathbf{y}$; in total energy balance, $\mathbf{y} = \mathbf{0}$. The vector $\mathbf{y}$ is directly computable from a complete description of the radiation field, the result of solving the transfer equation. All heating and cooling processes which are represented in either the emissivity, the absorption opacity, or the redistribution function---that is, bremsstrahlung, all atomic processes (photoionization, recombination, and line emission) and (inverse) Compton scattering---have their effects on the energy balance included in the expression for $\mathbf{y}$. Knowledge of $I_{\mu\varepsilon}$ in each cell constitutes a full description of the radiation field; similar to $\mathbf{y}$, we call such a collection $\mathbf{I}$. Similarly, the collection of energy-dependent absorption opacities and emissivities, $(\alpha_\varepsilon, j_\varepsilon)$, in all cells is denoted $\mathbf{S}$. Finally, the cell-by-cell list of temperatures is $\mathbf{T}$.

The first step in the procedure is to zero out all emission and absorption and perform a transfer solution with only \emph{Thomson} scattering. This yields a guess at the radiation field in each cell. For each cell independently, we supply to the relevant \textsc{xstar} subroutines the radiation field, density, temperature, and elemental abundances; \textsc{xstar} returns the photoionization equilibrium values for the ionization balance, emissivity (line and continuum), and absorption opacity. We do not use \textsc{xstar}'s built-in transfer apparatus. As in \citet{kin16a}, we ignore the resonant absorption of line photons on the basis of a \emph{post hoc} analysis of their escape probabilities, computed by \textsc{xstar}---due to the extremely high local turbulent velocity of the disk gas, these are all very near to unity. For this first iteration, we use \textsc{xstar}'s ability to find a local energy-conserving temperature while performing its photoionization equilibrium calculation. In addition to the heating and cooling rates \textsc{xstar} considers (see \citet{kal01a} for details), we also supply the \textsc{harm3d} simulation local cooling rate as an exogenous heating term. In subsequent iterations, we supply a local temperature according to the procedure described next. We now have a first guess at the absorption opacity, emissivity, and temperature in every cell. A second transfer solution performed with this $\mathbf{S}_0$ and $\mathbf{T}_0$ yields $\mathbf{I}_0$ and the corresponding $\mathbf{y}_0$.

We imagine our transfer/\textsc{xstar} scheme as a vector function: \textsc{xstar} requires $\mathbf{I}$ and $\mathbf{T}$ to determine the photoionization equilibrium $\mathbf{S}$, which via our transfer solution produces (a new) $\mathbf{I}$ and thus $\mathbf{y}$. That is, $F(\mathbf{I}, \mathbf{T}) = \mathbf{y}$. We seek a procedure by which, for a given $\mathbf{I}$, we can find the energy-conserving temperature structure $\mathbf{T}^*$, such that $F(\mathbf{I}, \mathbf{T^*}) = \mathbf{0}$.

To do so, we employ the multidimensional Newton-Raphson algorithm. Starting with $\mathbf{T}_0$ and a corresponding $\mathbf{y}_0$, we perform a finite difference estimation of a Jacobian of the form
\begin{equation}
J_{ij} = \frac{\partial y_i}{\partial T_j}.
\end{equation}
The new guess at the energy-conserving temperature is
\begin{equation}
\mathbf{T} = \mathbf{T}_0 - \mathrm{J}^{-1} \mathbf{y}_0.
\end{equation}
From this new temperature structure we determine the new $\mathbf{S}$ with \textsc{xstar}, and with that perform a transfer solution to find the new $\mathbf{y}$. We repeat the procedure until all elements of $\mathbf{y}$ are sufficiently close to zero, i.e., for all $i$, the two terms on the right-hand side of equation \ref{eq:y} differ by less than 1\%. Thus we find $\mathbf{T}^*$. As a practical matter, it occasionally happens that elements in the new temperature vector are not reasonable (for example, negative temperatures); this typically occurs in situations where there are relatively sharp changes in density or cooling rate. In these cases, we require an additional step before the next iteration: these ``problem'' cells are isolated and their individual $y$ roots found by varying only \emph{their own} $T$ using the secant or bisection methods; these new $T$ values replace their nonsensical counterparts in $\mathbf{T}$, and iteration resumes.

It is important to stress that at no step of the procedure described above is the radiation field supplied to \textsc{xstar} altered. After the energy-conserving temperature structure $\mathbf{T}^*$ is found, and the absorption opacities and emissivities everywhere re-computed with it, one last transfer solution gives us a new and by construction energy-conserving radiation field. In fact, the full procedure can be thought of as a function which takes some radiation field $\mathbf{I}$ as input and returns a new radiation field $\mathbf{I^*}$---this new radiation field is energy-conserving, but the gas is in photoionization equilibrium with the \emph{previous} radiation field $\mathbf{I}$. Naturally, then, we just repeat the entire process until $\mathbf{I^*} = \mathbf{I}$. Thus we accomplish our goal: we have a complete description of a radiation field for which energy is conserved everywhere and with which the gas is at all points in photoionization equilibrium.

In cases where the disk body is many Thomson depths in thickness, it becomes impractical (mainly due to memory constraints) to treat it as a single finite slab extending from one photosphere to the other. Rather, we are forced to set an interior boundary condition some number of Thomson depths inward from the photosphere (we choose 10, see section \ref{num_specs} for details); the natural choice is to assume a blackbody flux \emph{into} the computation volume from the otherwise excluded disk interior. This is similar to the interior boundary condition used by \citet{gar10a} (i.e., the radiative diffusion approximation \citep{ryb86a}), but we do not set the disk temperature in advance according to \citet{sha73a}. Like the temperature within the computation volume, this boundary temperature is not assumed \emph{a priori}. Rather, it is found in the exact same way as part of the same formalism. At the inner boundary, we define an additional element of $\mathbf{T}$, $T^{\text{bound}}$, and an additional element of $\mathbf{y}$,
\begin{equation}
y^{\text{bound}} = \pm \int_0^{\infty} d \varepsilon F_\varepsilon^{\text{bound}} - \frac{1}{2} \left( \mathrm{total}\ \mathcal{L}\ \mathrm{in\ disk\ interior} \right),
\end{equation}
with the upper sign corresponding to the upper disk layers and the lower sign to the lower disk. Defined so, $y^{\text{bound}} = 0$ (at both upper and lower interior boundaries) indicates that the net flux into the computation volume is equal to the total cooling rate \emph{excluded} by the computation volume---like all elements of $\mathbf{y} = \mathbf{0}$, it is a statement of energy conservation. With some reasonable starting guess for $T^{\text{bound}}$ (e.g., $2 \sigma T^4 = \mathrm{total}\ \mathcal{L}\ \mathrm{in\ disk\ interior}$), our multidimensional Newton-Raphson method will find the energy-conserving interior boundary temperature as part of its overall solution.

\subsection{The Reprocessed Spectrum}
\label{repr_spec}

Next we compute from the output of \textsc{ptransx} a \emph{new} seed photon spectrum at all $(r, \phi)$ points on the disk surface. This is done simply by performing one last radiative transfer solution but with the incident intensity set to zero---the seed photon spectrum includes only those photons \emph{emitted} by the gas in the disk, not those which are reflected by it (see section \ref{communication} below). In Figure \ref{fig:seed_flux} we compare the initially assumed hardened blackbody seed photon spectrum at one point on the disk surface to the \textsc{ptransx} output seed photon spectrum. The \textsc{ptransx} spectrum is broader, higher at all energies, has a prominent H-like Fe K$\alpha$ emission feature at 7 keV, and a small K-edge absorption dip near 9 keV. Though the \textsc{ptransx} spectrum is slightly harder, there is still virtually zero emission above 10 keV. This pattern generally holds (though with a variable dominant Fe ionization state) for all accretion rates we consider, and at all radii except for where the disk's total Thomson thickness is $\lesssim 1$---there the \textsc{ptransx} seed photon spectrum is just optically thin free-free emission. For the example point shown, the integrated power of the \textsc{ptransx} seed photons is $\sim$ 50\% greater than that of the hardened blackbody---this is because the \textsc{ptransx} seed photons must carry additional energy from Compton and photoionization heating of the disk due to its irradiation by the corona. \textsc{pandurata} then ray-traces these new seed photon packets from the disk surface with a limb-darkened angular distribution consistent with \textsc{ptransx}'s transfer solution. In the initial \textsc{pandurata} solution, however, photon packets which strike the disk surface are simply reflected; for this second pass, the absorption and energy redistribution of photons impinging the disk surface is informed by the \textsc{ptransx} output.

\begin{figure}
\plotone{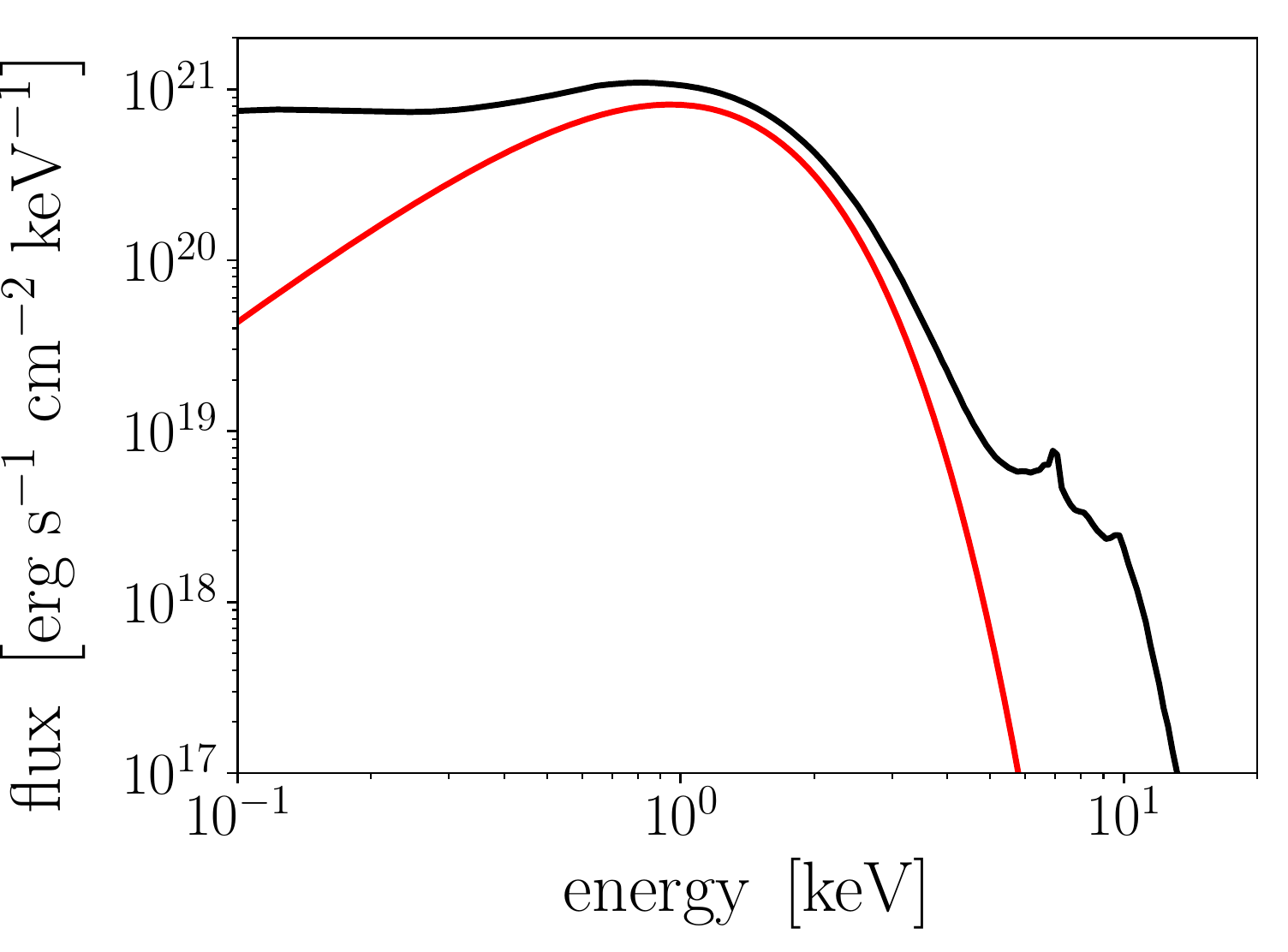}
\caption{Comparison between the initial assumed seed photon spectrum (red curve) and the converged \textsc{ptransx} output spectrum (black curve), at $r = 10M$, $\phi = 0$, for $\dot{m} = 0.03$. \label{fig:seed_flux}}
\end{figure}

When a photon packet with energy-dependent intensity $I_\varepsilon$ intersects the disk surface, it is transformed into a reflected photon packet with intensity $I_\varepsilon^{\text{refl}}$ according to
\begin{equation}
I_\varepsilon^{\text{refl}}/\varepsilon = \int_0^\infty d \varepsilon^\prime f \left( \varepsilon^\prime \right) G \left( \varepsilon, \varepsilon^\prime \right) I_{\varepsilon^\prime}/\varepsilon^\prime.
\label{eq:refl}
\end{equation}
The albedo $f$ is the fraction of incident photons at a specific energy which are reflected, i.e., not absorbed by the disk. $G$ is a normalized description of the redistribution of photons from energy $\varepsilon^\prime$ at incidence to energy $\varepsilon$ upon reflection. Both $f$ and $G$ are functions of position on the disk surface and of the photon packet's incident angle with respect to the local disk normal. These functions are tabulated using a separate, auxiliary Monte Carlo transport code. This additional code injects large numbers of photons at each energy and incident angle from the \textsc{ptransx} grids, for each point on the disk surface, using the \textsc{ptransx} output opacity and temperature structure; it records for each energy and angle the fraction which are reflected (the albedo $f$) and the energy-distribution of the reflected photons ($G$ in equation \ref{eq:refl}). The Compton scattering calculation therein is performed according to standard relativistic dynamics.

\begin{figure}
\plotone{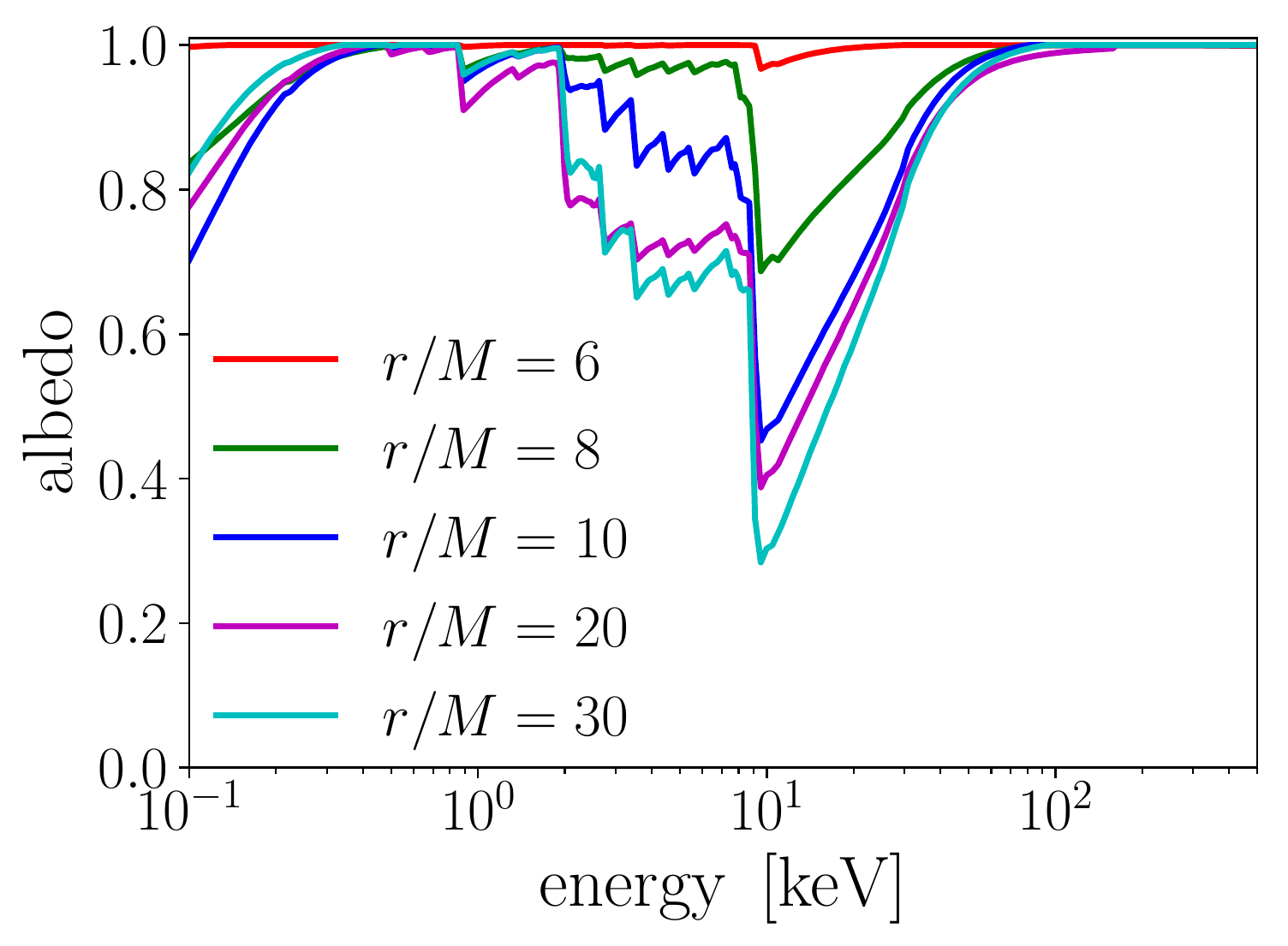}
\caption{The energy-dependent albedo (including transmitted fraction) at the disk surface at several radii (all $\phi = 0$) for $\dot{m} = 0.03$. \label{fig:albedo_rad}}
\end{figure}

\begin{figure}
\plotone{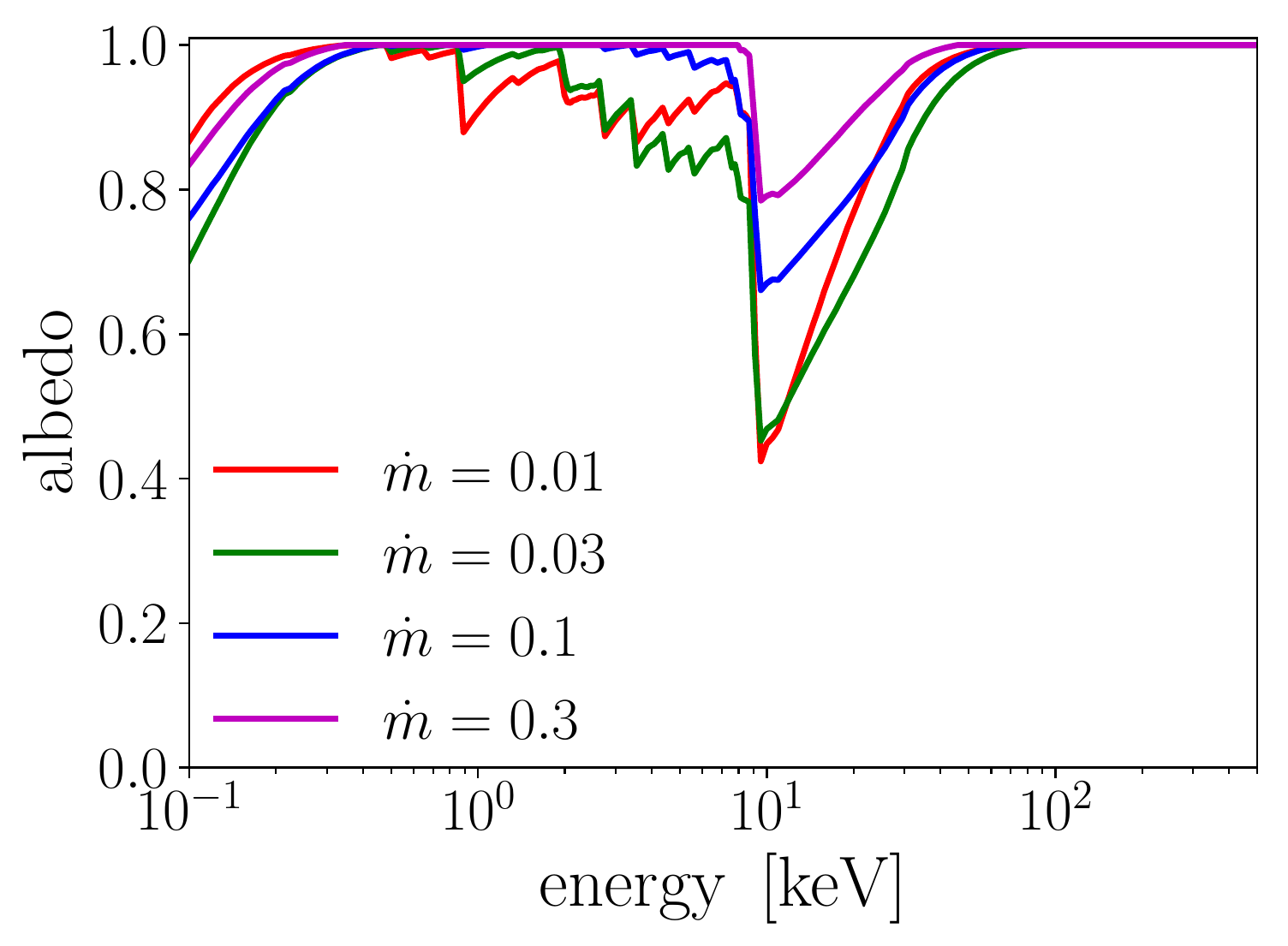}
\caption{The energy-dependent albedo (including transmitted fraction) at the disk surface at $r = 10 M$, $\phi = 0$, for several values of $\dot{m}$. \label{fig:albedo_mdot}}
\end{figure}

From this same Monte Carlo code, we have found that the distribution in angle of the outgoing photons is a very nearly linear function of $\mu$ with respect to the local disk normal (i.e., limb-darkening, but not exactly the pure scattering atmosphere expression of \citet{cha60a}); we therefore select the initial trajectory of the reflected photon packets according to this distribution. Small portions of the disk are only marginally optically thick, and in these regions there can be a significant \emph{transmitted} fraction. Ideally, the transmitted fraction would spawn a new photon packet in addition to the reflected packet. This is, however, not computationally feasible at this time, so instead we include the transmitted fraction in the reflected photon packet. To the extent that the upper and lower halves of the computation volume are similar, this will ultimately produce the same result. For the cases considered in this paper, transmission through the disk is negligible for $\gtrsim 99\%$ of the disk area. Figure \ref{fig:albedo_rad} shows the energy-dependent albedo at several radii for $\dot{m} = 0.03$; Figure \ref{fig:albedo_mdot} shows the albedo at $r = 10M$ for the four sample accretion rates. The most dramatic feature in both is, not surprisingly, the highly-ionized Fe K-edge at 8-10 keV: its depth increases at larger radii (as cooler, less-ionized gas has a higher fraction of unstripped Fe atoms available for absorption) and with decreasing $\dot{m}$ (for the same reason; see equations \ref{eq:rho} and \ref{eq:L} and Figure \ref{fig:albedo_mdot}).

\subsection{\textsc{ptransx} and \textsc{pandurata} Communication}
\label{communication}

The revised seed photon packets are the reprocessed emission from the disk, including atomic emission features like the Fe K lines. As \textsc{pandurata} transports them through the corona, they experience inverse Compton scattering in addition to all special and general relativistic effects. When they scatter off the disk surface, absorption features like the Fe K-edge are imprinted. A different spectrum of seed photons than that originally assumed affects the efficiency of the IC cooling process in the corona---so we run \textsc{pandurata} again, with disk albedo and Compton recoil tables in hand from the last \textsc{ptransx} run, and thus determine a new coronal temperature map and radiation field. This \textsc{pandurata} run yields a new irradiating flux, and so we run \textsc{ptransx} again to obtain new seed photon spectra and albedo tables. The cycle repeats until both disk and coronal radiation fields have converged.

Our method separates \emph{emission} from the disk and \emph{absorption}/\emph{reflection} by the disk into sequential steps: Fe K$\alpha$ line photons, for example, are emitted as part of the seed photon flux for a point on the disk surface consistent with the incident flux at that point found from the previous \textsc{pandurata} run; likewise, photon packets which strike the disk as they are ray-traced through the corona are subject to absorption according to the disk's opacity found in the preceding \textsc{ptransx} run. The same is true for the overall energy balance---the power in the seed photon flux accounts for both the disk's internal dissipation \emph{and} Compton and photoionization heating of the disk's gas consistent with the incident flux from, again, the previous \textsc{pandurata} run. By cycling between the two codes until the global radiation field (including the disk incident flux) no longer changes, we ensure \emph{global} energy balance and the self-consistency of our non-simultaneous treatment of absorption and emission.

The final run of \textsc{pandurata} yields the desired product: a spectrum as seen by a distant observer---a \emph{prediction}, arrived at through consideration of the relevant physical principles, for what we expect to observe from an accreting stellar-mass black hole, specifying only the \emph{physical} parameters of mass, spin, accretion rate, and elemental abundances.

\subsection{Numerical Specifics}
\label{num_specs}

In describing our technique, we have been intentionally vague concerning any numerical values. While our general approach is applicable to a large volume of the stellar-mass (and even AGN) black hole parameter space, the specific resolutions, samplings, etc., that we use in a real calculation must be tailored to the kind of problem we want to solve---we must balance the desired accuracy and completeness of our prediction with realistic computational constraints. In practice, this requires numerical experimentation: resolutions and samplings start off coarse and are repeatedly refined until the final results (presented in the next section) no longer appear to change.

Here we consider four cases, at $\dot{M}/\dot{M}_{\text{Edd}} = \dot{m} = 0.01,\ 0.03,\ 0.1,\ \mathrm{and}\ 0.3$; per equations \ref{eq:rho} and \ref{eq:L} in section \ref{sim_data}, the choice of $\dot{m}$ translates \textsc{harm3d} data snapshots into physical (cgs) values for the density and cooling rate. For each, the central black hole mass is $10 M_\odot$, the spin is zero, and the abundances are solar (\citet{gre96a} values). When running \textsc{pandurata}, the coronal volume is divided into $\sim 24,000$ sectors (the exact number varies with the location of the disk photosphere, and therefore decreases with increasing accretion rate) of $\Delta \theta = \pi/36$ and $\Delta \phi = \pi/32$ radians each, with a logarithmically increasing radial extent such that $\Delta r/r = 0.062$, starting at the event horizon, $r = 2M$. For the snapshots we used, smaller sectors than these do not result in an appreciable change to the shape or strength of the X-ray flux incident upon the disk surface or seen by the distant observer---doubling the number of sectors results in less than a 1\% change to the observable spectrum in the 1-30 keV range. The majority of these sectors lie wholly in the corona, but those which overlap with the disk have only their coronal part included in \textsc{pandurata}'s calculation. When running \textsc{ptransx}, we sample $\sim$ 300-500 $(r, \phi)$ points per case (with more for the higher accretion rate cases as the inner edge of the disk photosphere extends further inward). These are chosen uniformly in azimuth (at 8 $\phi$ angles) and logarithmically in $r$ such that $\Delta r/r = 0.062$. For each $(r, \phi)$ slab, the vertical cells are spaced semi-logarithmically in Thomson depth $\tau$, such that the increase in Thomson depth \emph{into} the disk over one cell is $\Delta \tau/\tau = 0.25$, but with $\Delta \tau$ limited to a maximum of 0.4; the grid is laid out so that the cells follow this semi-logarithmic spacing \emph{into} the disk starting from both upper and lower photospheres, meeting at the midplane. Slabs with a total Thomson depth of 20 or greater are cleaved into an upper and lower computation volume as described above, with the interior boundary always placed at a Thomson depth of 10 as measured from the relevant photosphere. The number of vertical cells used varies from 6 at the extreme inner edge of the disk to 50 at its thickest extent before the approximation just described is employed; the ``cleaved'' slabs are separated into upper and lower volumes of 27 cells each. Neither finer spacing in optical depth nor a deeper interior boundary result in an appreciable change to the output seed photon spectrum.

The angle with respect to $\hat{z}$, the cosine of which is the $\mu$ in our transfer equation, is discretized such that 16 bins uniformly spaced in $\mu$ cover the range -1 to 1. This is more than sufficient to capture the angular dependence of the radiation field, which is nearly isotropic for most of the disk body. Our energy grid is more complex. For the purposes of determining the temperature structure and photoionization balance via multidimensional Newton-Raphson, we span the range from 1 eV to $10^8$  eV with a coarse 161-point grid whose energy resolution is $\Delta \varepsilon/\varepsilon = 0.122$. The computational cost of the transfer solution scales poorly with the number of energy bins (cubically) and our multidimensional Newton-Raphson scheme requires the transfer solution to be performed \emph{many} times---typically 20-80 iterations per slab, depending on its thickness and ionization parameter. Yet because the bulk of the power in the radiation field is in the continuum, and the broadband continuum can be well-represented on such a coarse grid, increasing the energy resolution further results in little change in the equilibrium temperature structure. Thus the approach we take is to use a coarse grid to find the equilibrium temperature structure, then re-bin to a finer 801-point grid ($\Delta \varepsilon/\varepsilon = 0.0233$) on which we perform \emph{one} last transfer solution at a resolution high enough so that line features are clearly distinguishable; we use this same procedure (and identical energy grids) with \textsc{pandurata} as well.

We take 1\% as sufficient for all convergence tests. That is, ``energy conservation'' (for both \textsc{ptransx} and \textsc{pandurata}) means (energy in) = (energy out) is satisfied in all cells/sectors (and globally) to within at most 1\%. The majority are better converged by the time this is achieved---typically, $\sim 90\%$ of cells/sectors conserve energy within 0.1\%. For determining if $\mathbf{I^*} = \mathbf{I}$, we compute the first several energy moments of the mean photon intensity in the range 1-30 keV (the region of the outgoing spectrum we are most concerned with) at each cell; when the greatest fractional difference between any of these values and its counterpart in the previous iteration has dropped below $1\%$, we consider the radiation field to have converged. Finally, the cycling between \textsc{ptransx} and \textsc{pandurata} ceases when the greatest difference between the spectrum as seen by a distant observer  (at any inclination or energy in the range 1-30 keV) from one iteration to the next differs by, again, at most 1\%---this takes, for the cases we discuss in this paper, between 5-10 passes.

\section{Results}
\subsection{Continuum}

\begin{figure}
\plotone{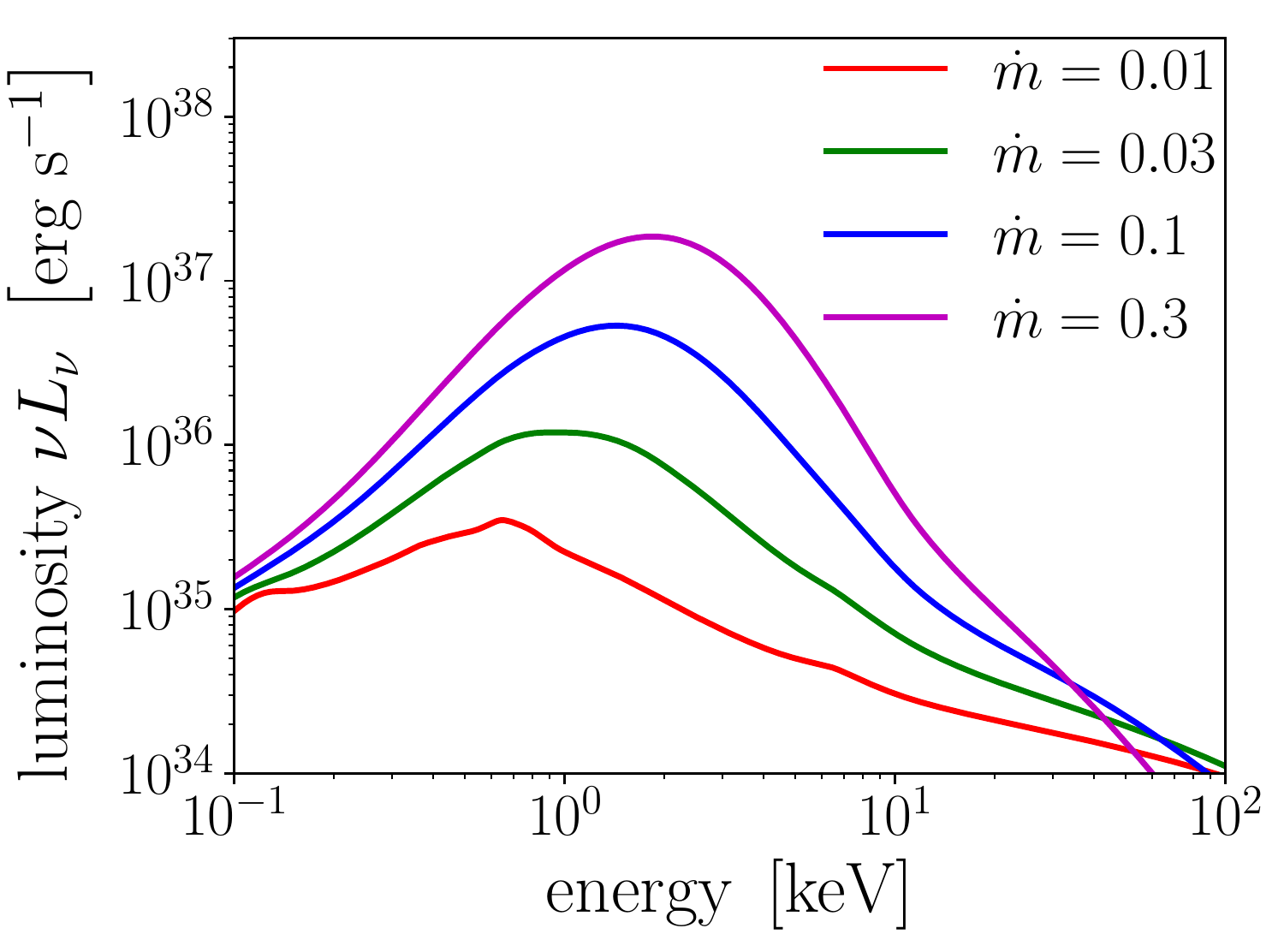}
\caption{The spectral luminosity at four accretion rates, each with a central black hole mass of $10 M_\odot$ and $a = 0$. Note that the spectra soften as the accretion rate increases. Although not easily visible in this representation, the equivalent width of the Fe K$\alpha$ feature diminishes with increasing $\dot{m}$. \label{fig:lum}}
\end{figure}

The key results of our calculation are predicted X-ray spectra as seen by a distant observer. Figure \ref{fig:lum} shows the broadband spectral luminosity for the four accretion rate cases we consider. Because the dynamical timescale for stellar-mass black holes is many times smaller than the integration time for any reasonable observation of them, we present all distant observer spectra as azimuthally-averaged. Particularly for the $\dot{m} \geq 0.03$ cases, these broadband spectra reproduce the forms inferred by phenomenological fitting of real black hole X-ray binary data in the steep power-law state \citep{rem06a}: there is a quasi-thermal bump at 1-3 keV that is extended to high energy as a steep power-law that hardens slightly above $\sim 10$ keV. The photon index $\Gamma$ computed over the 2-30 keV band ranges from 2.7 for $\dot{m} = 0.01$ to 4.5 for $\dot{m} = 0.3$, comparable to the observed values for black hole binaries in the thermal and steep power-law states, $\Gamma = 2.1$-4.8 \citep{mcc06a}. The thermal bump is not too surprising, and its existence and temperature follow from having a dense, optically thick disk body with a sub-Eddington accretion rate around a $\sim 10 M_\odot$ black hole. On the other hand, the steep power-law component represents a triumph for the theory of MHD accretion disks. No coronal emission at all is predicted by the models of \citet{sha73a} and \citet{nov73a}. We find here that a purely physical calculation, starting from a GRMHD simulation of black hole accretion, gives rise \emph{naturally} to the approximate spectral shape observed, with no phenomenological descriptions of the accretion geometry (of disk or corona) or parameter-tweaking required. This result was first shown by \citet{sch13a}, also using \textsc{pandurata} analysis of \textsc{harm3d} simulations. With our more careful treatment of the seed photon spectrum and the inclusion of disk absorption enabled by coupling to \textsc{ptransx}, we predict slightly softer spectra than those reported in \citet{sch13a}. Curiously, using simulations nominally similar to ours (GRMHD thin disk simulations with prescribed cooling functions) and a post-processing procedure calculating the Comptonization of initially thermal photons, \citet{nar16a} were unable to find any high-energy extension of the thermal component.

\begin{figure}
\plotone{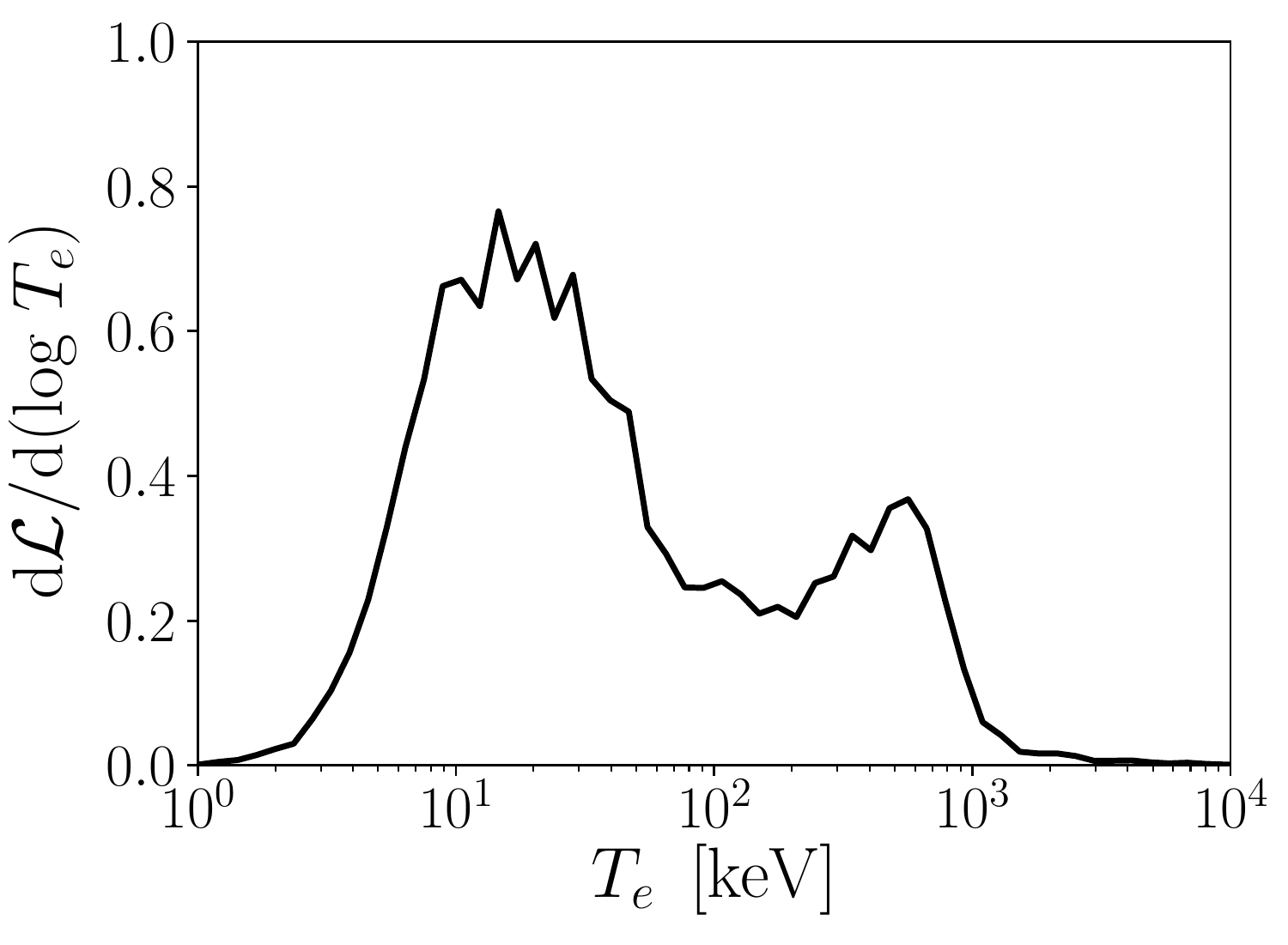}
\caption{The distribution of IC power as a function of electron temperature in the corona for $\dot{m} = 0.03$. The data are normalized to the total cooling in the corona. Note both the broad range in temperature and the distinctly bimodal shape of the distribution. \label{fig:dLdT}}
\end{figure}

The spectra in Figure \ref{fig:lum} have power-law tails which extend to very high energies. In Figure \ref{fig:dLdT}, we show (for $\dot{m} = 0.03$) the distribution of IC power generation in the corona as a function of the electron temperature. The photons which make up the observable spectrum were upscattered by electrons with a wide range of temperature, 1-1000 keV, but the distribution is distinctly bimodal with peaks at 10-30 keV and 400-800 keV. While the majority of the cooling is due to electrons with temperatures less than 100 keV, approximately 20\% of the coronal power is radiated from electrons with temperatures in excess of 400 keV. The other three accretion rate cases also have similarly bimodal distributions. It is evident from these figures that a single-temperature Comptonization model of the corona cannot adequately describe our results. Similarly, a single Compton $y$-parameter does not adequately describe coronal scattering. A thermal 1 keV seed photon which escapes to the distant observer will typically undergo 3-7 scatters. If it scatters through electrons at $T_e = 20$ keV (roughly the location of the first peak in Figure \ref{fig:dLdT}), then $y \approx 0.7$; however, if it scatters through electrons at $T_e = 500$ keV (roughly the location of the second, smaller peak), $y \approx 70$.

We do not see clear evidence for Compton bumps in the spectra of Figure \ref{fig:lum}. These bumps can be seen most clearly when the disk is absorptive across the soft X-ray band and up through the Fe K-edge, and when the underlying continuum is relatively hard so that there are plentiful photons above $\sim 10$ keV to scatter. Here neither is the case. The result is that the numerous photons with energy below the onset of absorption at the Fe K-edge can be upscattered and smooth over the feature. We expect the reflection hump to be visible when we scale these same simulations to AGN masses and temperatures. At that scale, we expect elements other than Fe to produce important spectral features as well.

\begin{figure}
\plotone{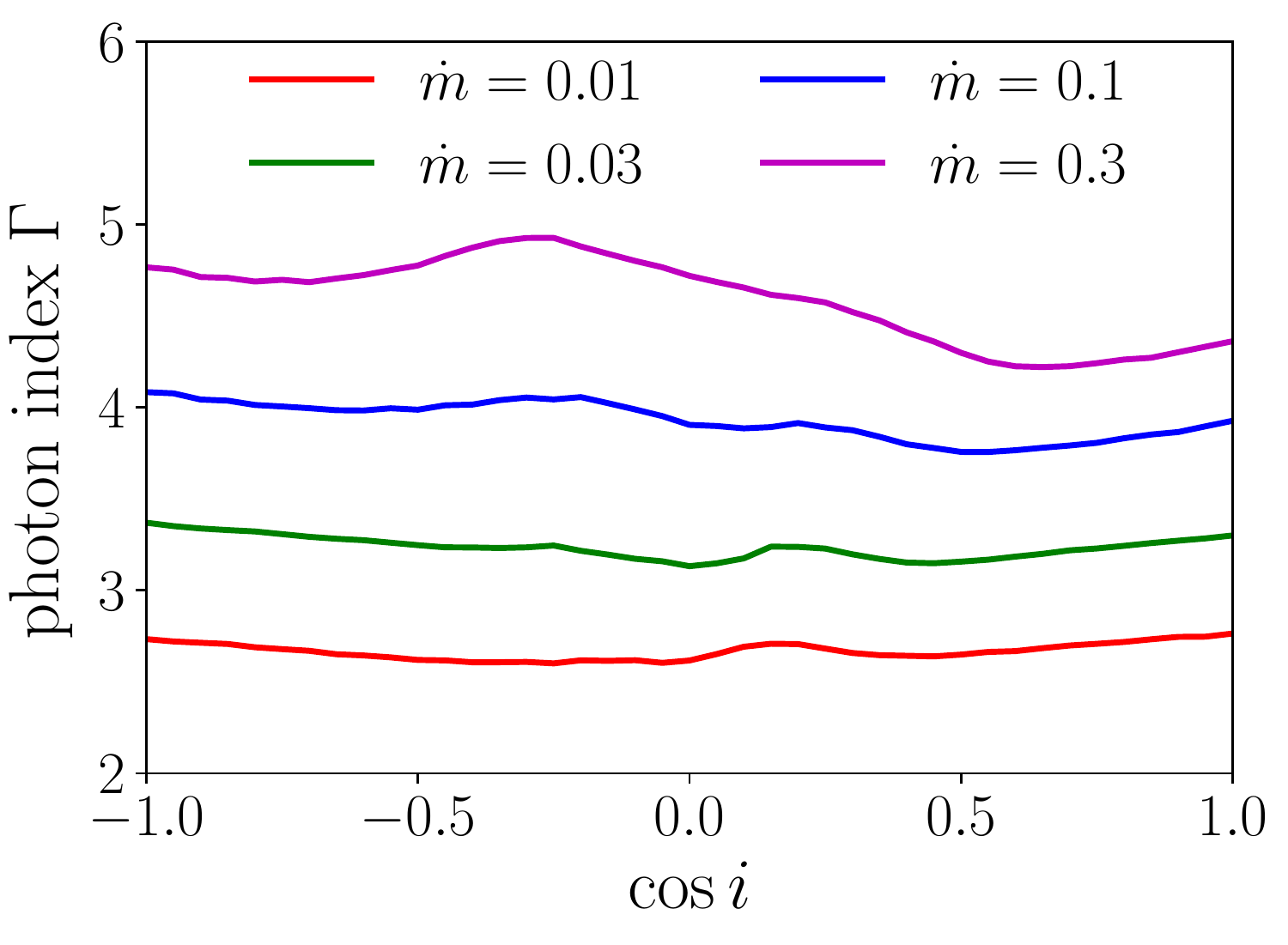}
\caption{The photon index of the predicted spectrum, measured in the range 2-30 keV, as a function of observer angle at four accretion rates. \label{fig:gamma}}
\end{figure}

Figure \ref{fig:gamma} shows the photon index measured in the range 2-30 keV as a function of observer angle for each accretion rate. The power-law slope varies irregularly, but only slightly, with inclination; its range increases with increasing $\dot{m}$. The extent of the top-bottom asymmetry seen in Figure \ref{fig:gamma} (and also in Figure \ref{fig:ew} below, showing Fe K$\alpha$ equivalent width in the same fashion) provides a rough indication of the ``cosmic variance'' expected for these simulations.

\pagebreak

\subsection{Fe K$\alpha$}

\begin{figure}
\plotone{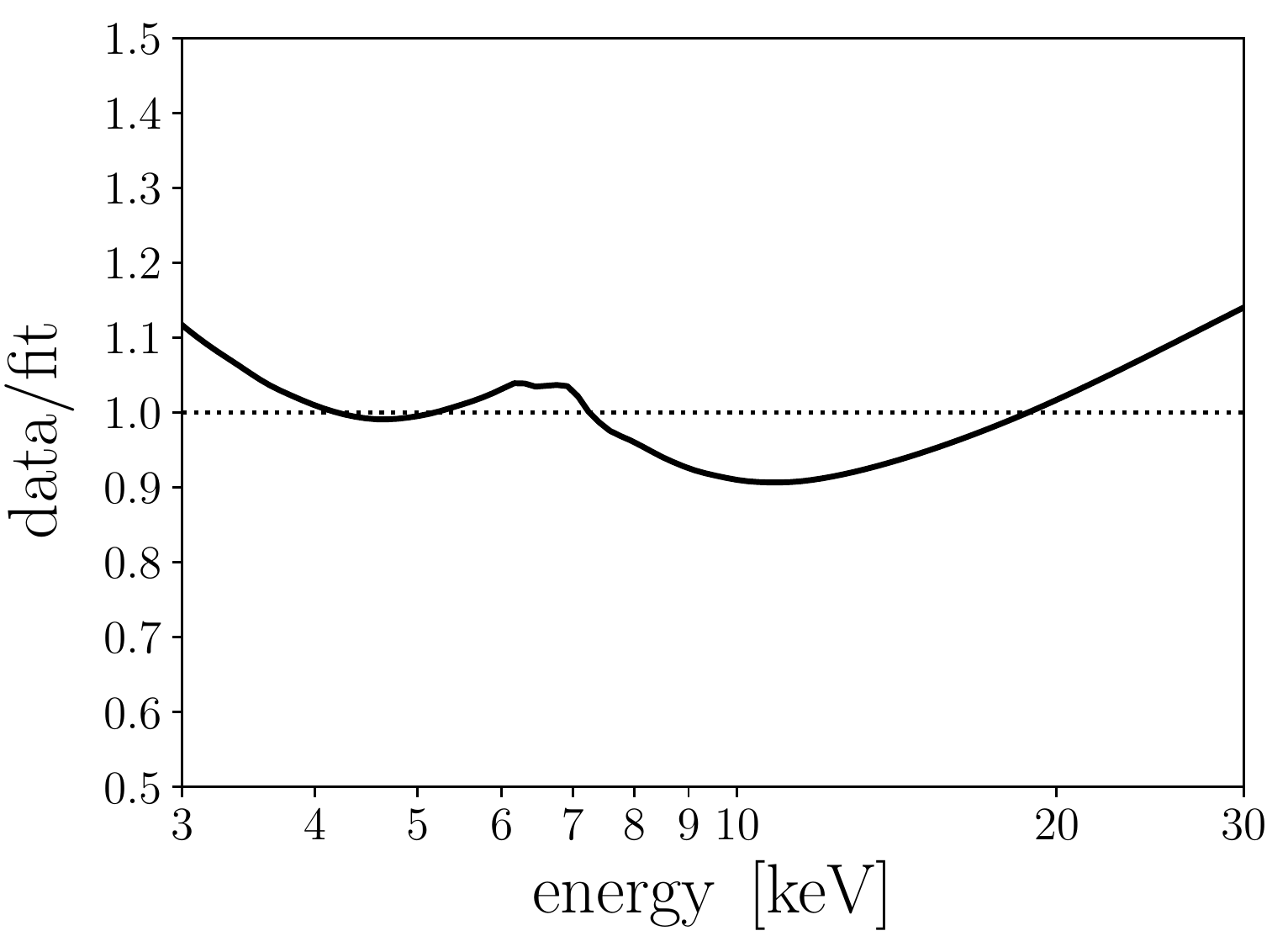}
\caption{The predicted spectrum divided by a power-law fit to the range shown for $\dot{m} = 0.01$ and $i = 25^\circ$. \label{fig:pl_ratio}}
\end{figure}

To illustrate our predictions of the Fe K$\alpha$ line profile, we adopt a procedure mimicking a common approach to presenting observational data: we divide the data by a simple prescription for the continuum---in this case, a power-law fit to the region 3-30 keV. Figure \ref{fig:pl_ratio} shows this procedure applied to the $\dot{m} = 0.01$ case at an observer inclination of $25^\circ$, for which the fitted power-law has photon index $\Gamma = 2.7$. We reproduce the features expected: a relativistically-broadened K$\alpha$ emission line near 6.4 keV and a K-edge absorption trough centered roughly at 10 keV. However, the contrast of both features relative to the power-law fit is quite small, only 5-10\%. In addition, above 15-20 keV there is a slight hardening of the continuum relative to $\Gamma = 2.7$, the single value that best fits the 3-30 keV continuum slope.

\begin{figure}
\plotone{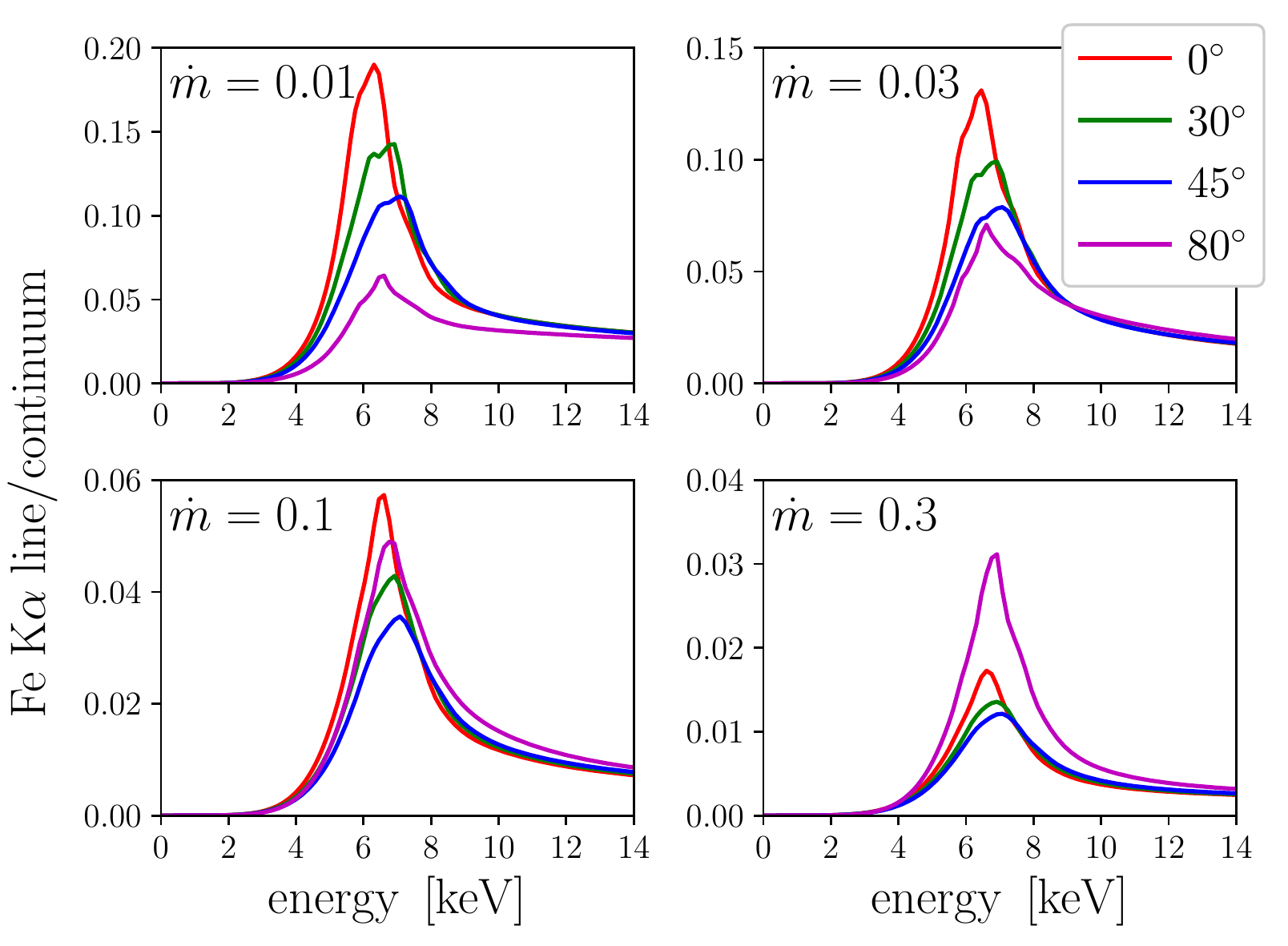}
\caption{Photons whose origin is an Fe K$\alpha$ transition, as a fraction of all continuum photons, once they have reached the distant observer; for four accretion rate cases at several inclinations each. The ``shelf'' at high energy is due to IC upscattering in the corona taken in ratio to a steeply-declining continuum. Note the difference in scale for each subplot. \label{fig:lines}}
\end{figure}

\begin{figure}
\plotone{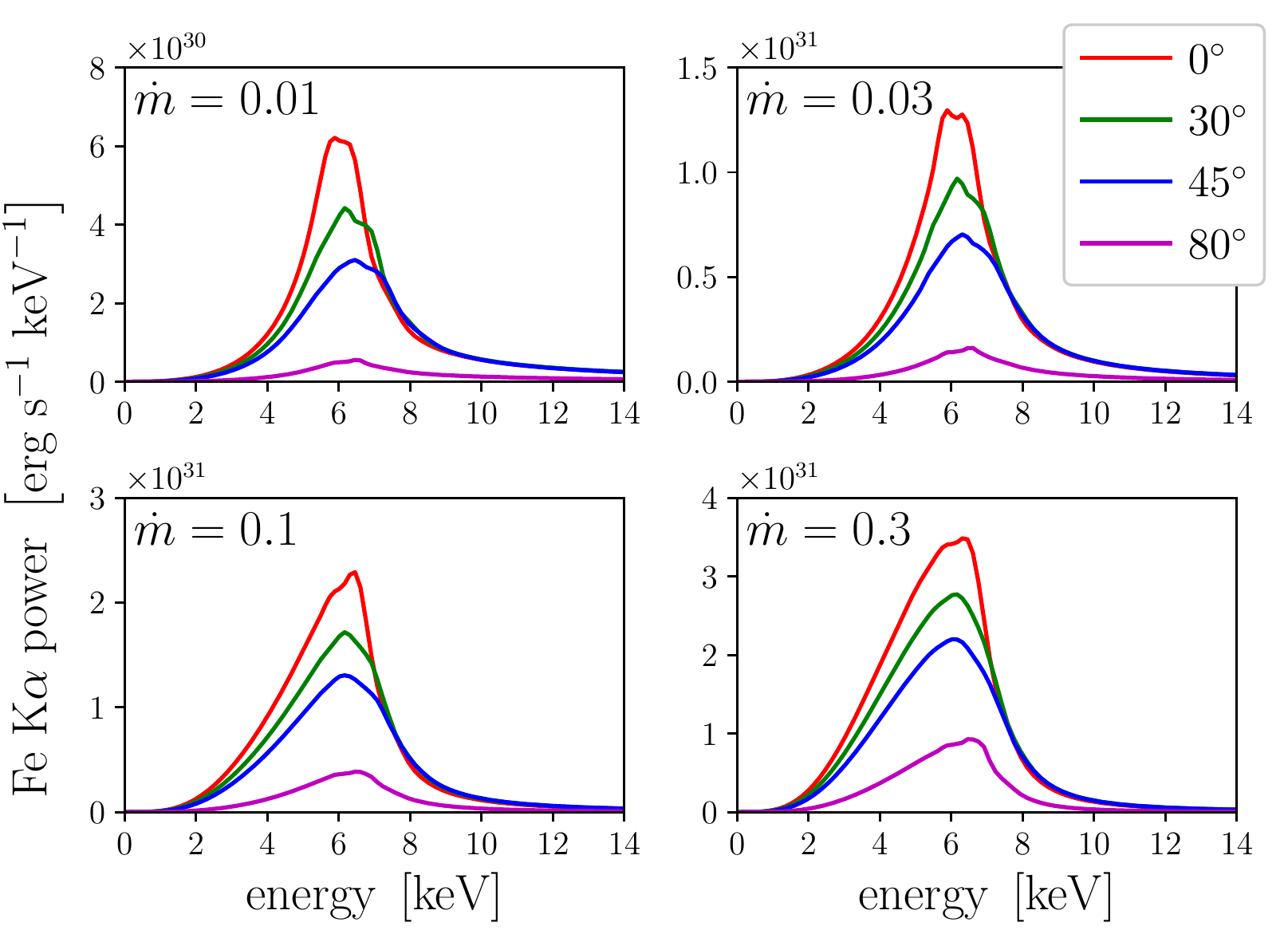}
\caption{The Fe K$\alpha$ spectral flux, for four accretion rate cases at several inclinations each. Note that the high energy ``shelf'' is considerably more modest in this representation than it appears in Figure \ref{fig:lines}. Note also the difference in scale for each subplot. \label{fig:line_fluxes}}
\end{figure}

\begin{figure}
\plotone{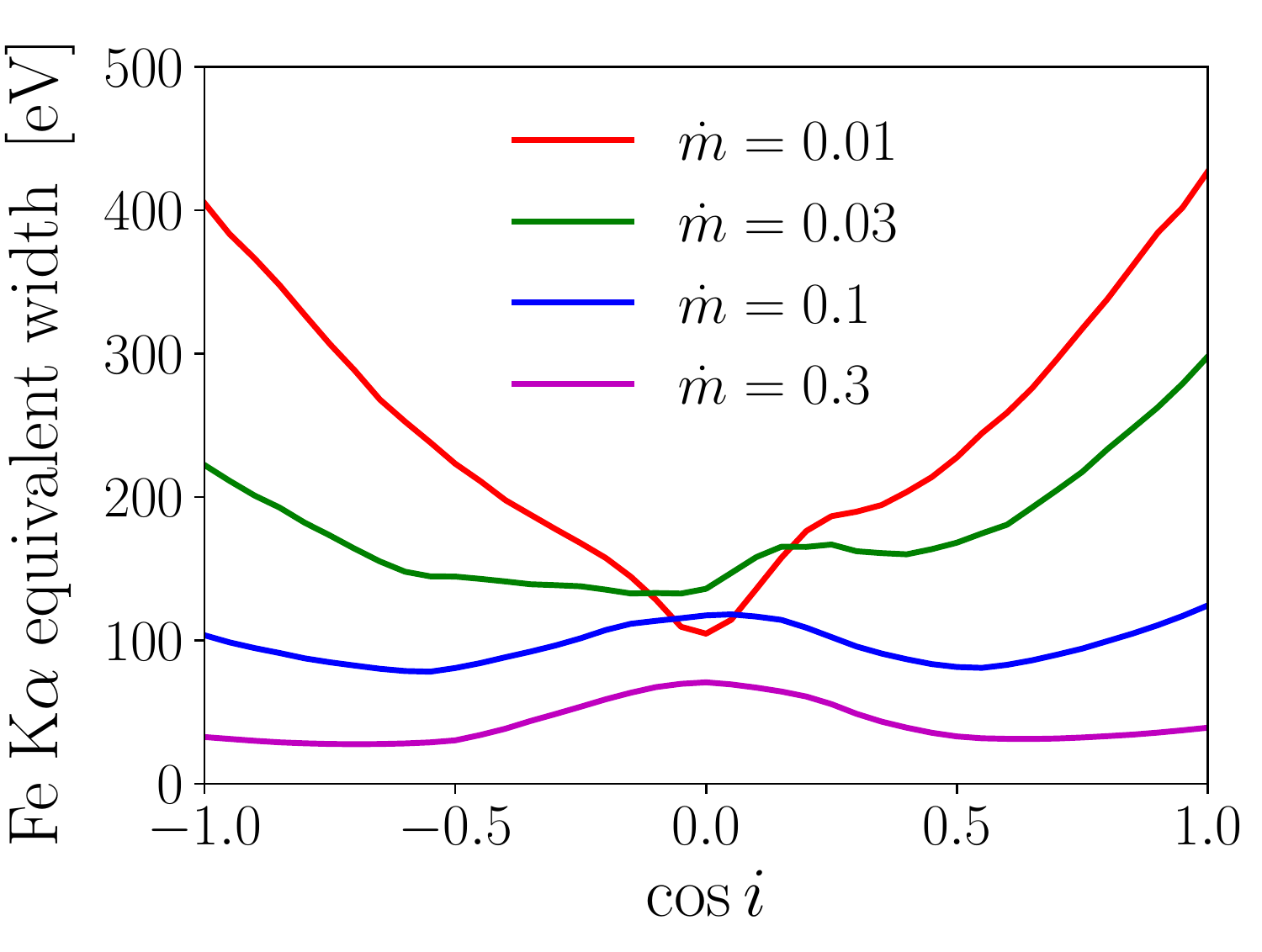}
\caption{The Fe K$\alpha$ equivalent width, as a function of observer angle, for each of the four accretion rate cases we consider. The EW is computed in the 2-8 keV range. \label{fig:ew}}
\end{figure}

In Figure \ref{fig:lines}, we show those photons which \emph{originate} from the K$\alpha$ transition as a fraction of continuum photons at the same energy, as seen by a distant observer, at several inclinations for each accretion rate. It is important to note that while this representation emulates model-fitting procedures, these are \emph{not} themselves model fits divided by the total flux. We produce these plots by keeping track of K$\alpha$ photons as they diffuse from their point of creation to the disk surface and are then ray-traced to infinity, with no continuum-fitting needed. In Figure \ref{fig:line_fluxes} we show the line fluxes \emph{only} (continuum-subtraction rather than continuum-division). We calculate the equivalent width (EW) directly as well, as shown in Figure \ref{fig:ew} as a function of observer angle.

The Fe K$\alpha$ line profiles strongly resemble actual spectral data in the sense that they are fairly broad and their EWs are in the range often measured ($\sim 100$ eV, see below). On the other hand, they also differ in some respects. In particular, the ``shelf'' at high energies in Figure \ref{fig:lines} is due to the upscattering of Fe K$\alpha$ photons as they traverse the corona. Note that this ``shelf'' line flux is \emph{not} accounted for by continuum-fitting models; rather, these upscattered Fe K$\alpha$ photons would appear as continuum photons, potentially introducing a systematic bias in a continuum-subtraction procedure as the level of the perceived continuum at energies above the K$\alpha$ feature is artificially elevated by a few percent. This coronal upscattering effect, in addition to our use of a non-spinning black hole, also explains the blueward asymmetry of the line profiles in Figure \ref{fig:lines}. In the flux-subtracted representation of Figure \ref{fig:line_fluxes}, the line profiles are more symmetric, as the apparent strength of the blue wing is enhanced in the representation of Figure \ref{fig:lines} due to the division by the underlying steep-power law continuum. The simple data/fit representation of Figure \ref{fig:pl_ratio} is the closest approximation to actual observed line profiles: compared to, e.g., Cyg X-1 line profiles \citep{rey03a, wal16a}, ours are nearly as broad, though with a slightly less extended red wing.

The Fe K$\alpha$ feature becomes relatively weaker with respect to the continuum as the accretion rate increases---the peak contrast drops by a factor of six from $\dot{m} = 0.01$ to $\dot{m} = 0.3$ in Figure \ref{fig:lines}. This is consistent with the increasing ionization parameter \citep{tar69a}: $\log \xi$ varies with position (decreasing with radius), but is in the range 2.5-3.0 for $\dot{m} = 0.01$, increasing to 3.0-3.5, 3.5-4.0, and 4.0-4.5, for $\dot{m} = 0.03$, 0.1, and 0.3, respectively.

\begin{figure}
\plotone{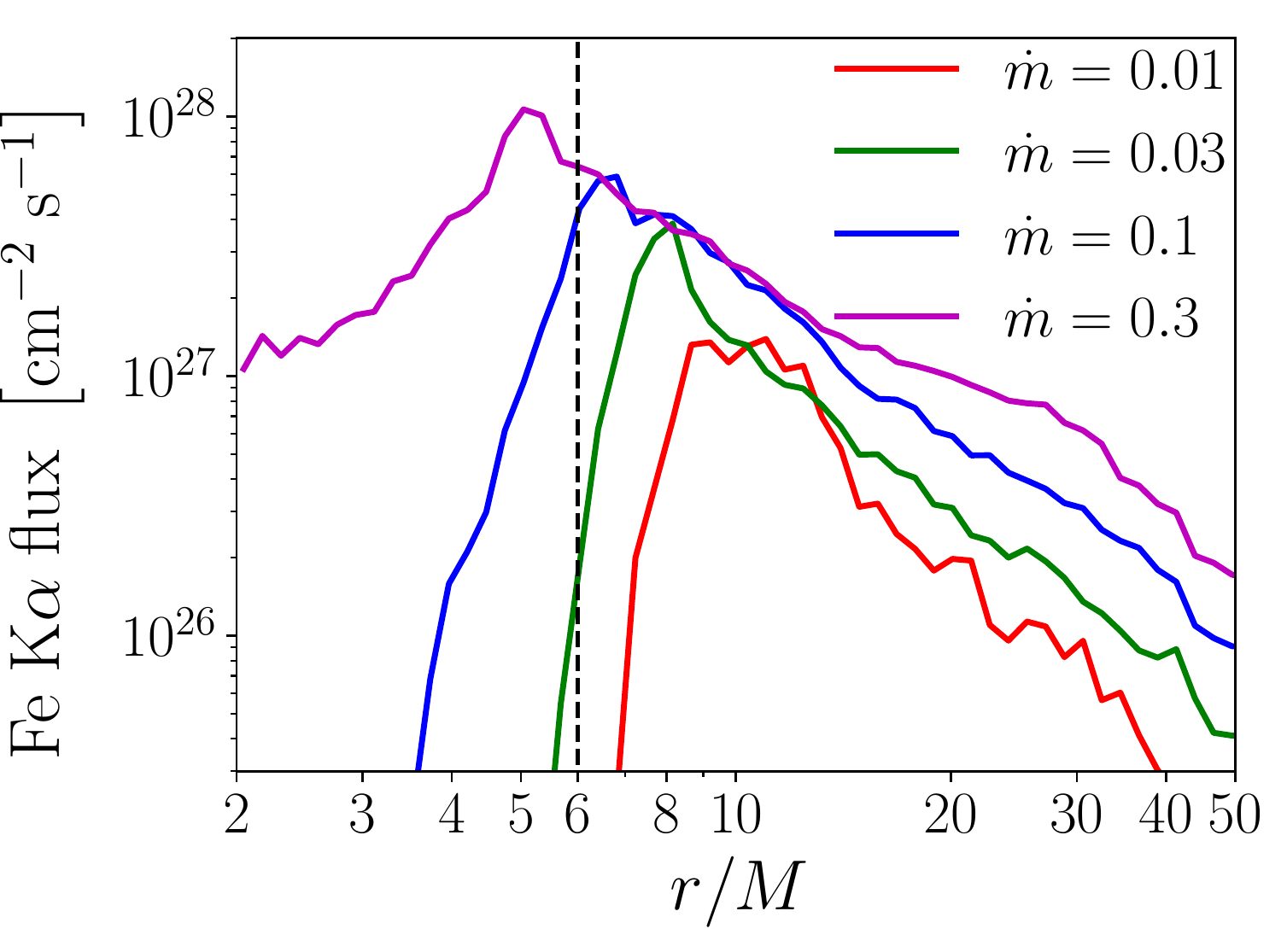}
\caption{Fe K$\alpha$ surface brightness, averaged over azimuth and both disk surfaces, for four accretion rates. Note the location of the peaks with respect to the ISCO at $r = 6M$. \label{fig:feka_rad}}
\end{figure}

For the lower accretion rates, the line contrast drops monotonically with increasing inclination. However, as $\dot{m}$ increases, the near edge-on ($i = 80^\circ$) view becomes stronger relative to the other viewing angles, and is twice the strength of the other three sample inclinations for $\dot{m} = 0.3$. Note that this is not obvious from the line flux plots of Figure \ref{fig:line_fluxes}. The overall line flux for the near edge-on viewing angles is less than the face-on inclinations for all accretion rates, but the strength of the line \emph{relative} to the continuum is greater when viewed near edge-on because the disk itself obscures emission from outer radii (where the line flux is weak compared to the continuum) while, due to lensing, light from the inner disk region (where the line flux is strong compared to the continuum) still reaches the distant observer; this is only the case when there \emph{is} any K$\alpha$ emission at the innermost radii, i.e., for higher accretion rates (see Figure \ref{fig:feka_rad}).

The line strengths we find are, overall, comparable to those typically observed, with EWs in the range 40-400 eV. Compare, for example, to the \citet{wal16a} analysis of Cyg X-1 soft state spectra; they report an Fe K$\alpha$ EW = 300-330 eV (see also the discussion in \citet{rey03a}). It is a well-known phenomenon that modeling-fitting of black hole X-ray spectra often results in inferred Fe abundances that are several-to-many times the solar value \citep{tom18a, gar18a}. Because we have not yet fit real data with our simulated spectra, nor do we analyze our theoretical spectra with the same techniques used by observers, it is not possible to make \emph{direct} comparisons between our line strengths and those reported in, e.g., \citet{wal16a}. Nevertheless, that we use \emph{only} solar Fe abundances yet still find strong K$\alpha$ lines is encouraging. Our approach is fundamentally different from those often employed when fitting actual spectra, so it is difficult to pinpoint a single reason why we do not need supersolar Fe abundances to achieve strong lines. A major contributor, however, is likely our naturally extended corona (see Figure \ref{fig:t_map}), which allows for Fe K$\alpha$ production over a larger fraction of the disk surface. Figure \ref{fig:feka_rad} shows the radial dependence of the Fe K$\alpha$ surface brightness for each accretion rate case, averaged over azimuth and the two surfaces of the disk. Like most phenomenological models, the variation with radius is roughly power-law---our power-laws, however, are approximately $\propto r^{-2}$ (steepening slightly with decreasing accretion rate), a shallower profile than is typically assumed from lamppost geometries ($\propto r^{-3}$ or steeper \citep{wil12a, dau13a}). Azimuthal variations superimposed on these radial gradients can be sizable: the relative standard deviation of the $\phi$-variation of the K$\alpha$ emission decreases with increasing $\dot{m}$, from $\sim 50\%$ for $\dot{m} = 0.01$ to $\lesssim 10\%$ for $\dot{m} = 0.3$.

The radius of peak K$\alpha$ surface brightness moves inward with increasing accretion rate, from $\sim 10 M$ for $\dot{m} = 0.01$ to $\sim 5 M$ for $\dot{m} = 0.3$. The K$\alpha$ surface brightness at radii interior to the peak increases with accretion rate as well; for $\dot{m} = 0.3$, there is significant K$\alpha$ production even just outside the event horizon. While the strength of the Fe line relative to the continuum diminishes with increasing accretion rate (Figure \ref{fig:ew}), the \emph{number} of K$\alpha$ photons increases as the accretion rate, and therefore the total luminosity, increases (compare the $y$-axis scales of the subplots in Figure \ref{fig:line_fluxes}). Though we do find that the peak in K$\alpha$ surface brightness occurs somewhat \emph{near} the ISCO, we do not find a sharp cutoff exactly \emph{at} the ISCO. This should not be too surprising. If gas flows into the black hole, there must be some gas between the event horizon and the ISCO which might produce Fe K$\alpha$ emission; conversely, if the accretion rate is low and the illuminating flux particularly powerful, there could be no available unstripped Fe to emit photons except for well outside the ISCO. In general, the interior K$\alpha$ emission cutoff cannot just be a function of the spin alone---it depends also on the disk's surface density and ionization state, which themselves depend on the accretion rate, as we demonstrate in Figure \ref{fig:feka_rad} (see also the discussions in \citet{rey97a}, \citet{kro02a}, and \citet{bec08b}).

We have considered only the $a = 0$ case here, but the variation of the peak surface brightness with accretion rate for \emph{non}-spinning black holes has important implications for spin-measuring techniques which rely on identifying the interior cutoff of K$\alpha$ production as the ISCO. In future work, we will apply our method to simulations with $a > 0$ as well. We have already done this for the continuum flux method for spin measurement \citep{sch16a}; it will be very interesting to see how the Fe line technique compares to the continuum method.

\pagebreak

\section{Conclusion}

The most important result is simply that our machinery can manufacture \emph{forward} predictions of the entire X-ray spectrum radiated by an accreting stellar-mass black hole---line \emph{and} continuum features together---in a self-consistent, energy-conserving fashion, directly from the output of high-resolution 3D GRMHD simulations. It is worth repeating that we have required \emph{no} assumptions about the accretion flow geometry at any point---no lamppost coronae, or disk inner edges fixed exactly at the ISCO. And yet, by specifying \emph{only} the physically meaningful parameters of mass, spin, accretion rate, and elemental abundances, and then applying to the simulation data the relevant physical principles and carrying out detailed radiative transfer and photoionization calculations, we are able to produce spectra similar in shape and principal features to those actually observed from stellar-mass black holes. It also bears emphasizing that we employed standard techniques without preference for a desired outcome---a Monte Carlo radiation transport code which treats only Compton scattering is the natural choice in the hot, optically thin corona, while a plane-parallel Feautrier method which treats Compton scattering, free-free, and all atomic processes, is the natural choice in the cooler, optically thick disk. That the output of these methods when applied to simulation data resemble so well the familiar X-ray spectra of their real counterparts encourages us to attempt to understand these objects from the standpoint of direct application of well-understood physical principles, as opposed to phenomenological modeling. In this vein, we ultimately plan to use this method for the production of grids of spectra---allowing observers to attempt to fit real spectral data with an \textsc{xspec} package that requires only a relatively small set of physical parameters. Moreover, as simulation codes are improved (in particular with regard to the equation of state), our methods can be readily employed upon the data they produce.

\acknowledgments

We thank Jon Miller, Andy Fabian, and Javier Garc\'{i}a for helpful conversations. This work was partially supported by NASA ATP grant NNX14AB43G and NSF grants AST-1516299 and AST-1715032. JDS was partially supported by NASA grant ATP13-0077. We are also grateful for generous allocations computer time through the XSEDE program of NSF (TG-MCA95C003) and at the Maryland Advanced Research Computing Center (MARCC).

\bibliography{references}

\end{document}